\documentclass[%
    aip,
     amsmath,amssymb,
     reprint,%
]{revtex4-1}
\usepackage{amsthm}
\usepackage{graphicx}
\usepackage{dcolumn}
\usepackage{bm}

\usepackage{dcolumn}

\usepackage[utf8]{inputenc}
\usepackage[T1]{fontenc}
\usepackage{mathptmx}
\usepackage{etoolbox}

\usepackage{soul}

\makeatletter
\def\@email#1#2{%
 \endgroup
 \patchcmd{\titleblock@produce}
  {\frontmatter@RRAPformat}
  {\frontmatter@RRAPformat{\produce@RRAP{*#1\href{mailto:#2}{#2}}}\frontmatter@RRAPformat}
  {}{}
}%
\makeatother

\usepackage[margin=2cm]{geometry}
\allowdisplaybreaks

\usepackage{subfig}

\usepackage{hyperref}
\usepackage[dvipsnames]{xcolor}
\usepackage{amsmath}
\usepackage{amsfonts,amssymb}

\usepackage{tikz}
\usetikzlibrary{quantikz2}

\usepackage{soul}  
\usepackage{pgfplots}
\pgfplotsset{compat=1.18}

\usepackage{mathbbol}
\usepackage{xifthen}

\usepackage[titletoc,title]{appendix}

\usepackage{multirow}

\colorlet{ygray}{gray!20}

\newcommand{\aeq}{\begin{equation}}
\newcommand{\eeq}{\end{equation}}
\newcommand{\aeqn}{\begin{eqnarray}}
\newcommand{\eeqn}{\end{eqnarray}}
\newcommand{\aeqns}{\begin{eqnarray*}}
\newcommand{\eeqns}{\end{eqnarray*}}

\newcommand{\yi}{\mathrm{i}}

\newcommand{\ycb}[1]{\textcolor{Black}{#1} }
\newcommand{\ycr}[1]{\textcolor{Black}{#1} }

\newcommand{\oO}{\mathcal{O}}

\newcommand*\diff{\mathop{}\!\mathrm{d}}

\DeclarePairedDelimiter{\yceil}{\lceil}{\rceil}

\newcommand{\yEx}[1][]{\ifthenelse{\isempty{#1}}{\tilde{E}_x}{\tilde{E}_{x,#1}}}
\newcommand{\yEy}[1][]{\ifthenelse{\isempty{#1}}{\tilde{E}_y}{\tilde{E}_{y,#1}}}
\newcommand{\yBz}[1][]{\ifthenelse{\isempty{#1}}{\tilde{B}_z}{\tilde{B}_{z,#1}}}

\def\kmax{k_{\rm max}}

\def\nk{n_k}
\def\nq{n_q}

\def\Nk{N_k}
\def\Nq{N_q}

\def\Ntrunc{N_{\rm trunc}}

\def\Reals{\mathbb{R}}

\newcommand{\abs}[1]{\left|#1\right|}
\newcommand{\norm}[1]{\left\|#1\right\|}

\newcommand{\yOw}{O_{\sqrt{w}}^{\rm AA}[U_{\sqrt{w}}]}
\newcommand{\yOwc}{\left(O_{\sqrt{w}}^{\rm AA}[U_{\sqrt{w}}^\dagger]\right)^\dagger}

\newcommand{\ybCm}{\|\bar{C}_{\rm max}\|}

\newcommand{\yerr}{\varepsilon_{\rm LCHS}}

\def\ktrunc{k_{\rm trunc}}
\def\kmax{k_{\rm max}}
\def\kmin{k_{\rm min}}

\usepgfplotslibrary{fillbetween}

\newtheorem{theorem}{Theorem}

\newtheorem{lemma}[theorem]{Lemma}

\begin{document}
\preprint{AIP/123-QED}

\title[]{An efficient explicit implementation of a near-optimal 
quantum algorithm for simulating linear dissipative differential equations}
\author{I. Novikau}
\affiliation{Lawrence Livermore National Laboratory, Livermore, California 94550, USA}
\author{I. Joseph}
\email{joseph5@llnl.gov}
\affiliation{Lawrence Livermore National Laboratory, Livermore, California 94550, USA}

\date{\today}

\begin{abstract}
We propose an efficient block-encoding technique for the implementation of the Linear Combination of Hamiltonian Simulations (LCHS) for \ycr{simulating} 
dissipative initial-value problems.
This algorithm approximates a target nonunitary operator as a weighted sum of Hamiltonian evolutions, thereby emulating a dissipative problem by mixing various time scales.
We introduce an efficient encoding of the LCHS into a quantum circuit based on a simple coordinate transformation that turns the dependence on the summation index into a trigonometric function.
\ycb{Classically, this method is equivalent to the use of a highly accurate Fej\'er-Clenshaw-Curtis quadrature formula.
Quantumly,} this significantly simplifies block-encoding of a dissipative problem and allows one to perform an exponential number of Hamiltonian simulations by a single Quantum Signal Processing (QSP) circuit.
The resulting LCHS circuit has high success probability and \ycb{the selector} scales logarithmically with the number of terms in the LCHS sum and linearly with time.
\ycb{Careful analysis of error convergence proves that this method is more efficient than other LCHS circuits that have recently appeared in the literature.}
We verify the quantum circuit and its scaling by simulating it on a digital emulator of fault-tolerant quantum computers and, as a test problem, solve the advection-diffusion equation.
The proposed algorithm can be used for modeling a wide class of nonunitary initial-value problems including the Liouville equation \ycb{with added dissipation} and linear embeddings of nonlinear systems\ycb{, such as the Koopman-von Neumann and Carleman embeddings.}
\end{abstract}
\maketitle


\section{Introduction}\label{sec:introduction}
\subsection{\ycb{Motivation}}\label{sec:motivation}
Precise modeling of complex dynamics characterized by multiple scales in space and time needs high spatial resolution.
As a result, accurate simulations of many classical problems of practical interest require large amounts of numerical resources.
This significantly limits the number of problems that can be simulated on classical computers, and it seems suitable to consider quantum computers (QCs) as possible candidates for modeling these sorts of problems.
By leveraging quantum effects such as entanglement and superposition of quantum states, QCs can process exponentially many complex numbers in parallel.
Moreover, in advanced quantum algorithms for solving differential equations, the number of operations grows polynomially or even exponentially more slowly than in the corresponding classical methods.
Yet, quantum computers can operate only with linear unitary operators that makes quantum simulations of dissipative problems challenging.

To solve dissipative differential equations, one can transform them into a system of linear equations\cite{Krovi23} $M\psi = b$ with a dilated matrix $M$.
The scaling of the matrix condition number $\kappa_M$ on the integration time depends on the Courant-Friedrichs-Lewy convergence condition and, for instance, is quadratic for the advection-diffusion or heat equations,\cite{Linden22} which have second-order spatial derivatives. 
The linear system can be solved by using a broad variety of efficient Quantum Linear System Algorithms (QLSAs) such as those based on variable-time amplitude amplification
\cite{Ambainis12, Childs17, Chakraborty19} or discrete adiabatic theorem,\cite{An22QLSA-adiab, Costa21} whose query complexity is linear with the condition number. 
More precisely, the number of calls to the oracle encoding $M$ and to the oracle encoding the right-hand-side vector $b$ is linear with $\kappa_M$.

Another option is the time-marching (TM) methods where one of the key issues is the decreasing success probability in time.
In Ref.~\onlinecite{Fang23}, it was proposed to use a combination of the uniform singular value amplification and compression gadgets to guarantee high success probability.
The resulting algorithm scales quadratically with the simulated time.
In Ref.~\onlinecite{Over24}, it was shown how to solve the advection-diffusion equation with the TM technique having linear complexity with time and high success probability.
However, it is not clear whether this method would preserve the near-optimal scaling if the normalized advection-like component of the TM matrix cannot be block-encoded keeping the spectral norm of the encoded matrix equal to one.
Also, explicit and implicit hybrid classical-quantum TM methods were proposed in Ref.~\onlinecite{Bharadwaj24}.
The TM algorithm was also used for modeling the Carleman linearization of the advection-diffusion-reaction equation.\cite{Sanavio24}

One can also use hybrid classical-quantum methods for solving dissipative problems on quantum computers as proposed, for instance, in Ref.~\onlinecite{Demirdjian22, Ingelmann24}.
Yet, the hybrid methods usually suffer from the bottleneck related to the repetitive input/output (I/O) data transfer between classical and quantum processors. 

An alternative approach is based on recasting a nonunitary operator associated with dissipative dynamics into a combination of unitary evolution operators integrated over \ycb{an} 
additional Fourier space.
This technique was initially proposed in Refs.~\onlinecite{Jin22Sch, Jin23Sch, Hu24Sch, Lu24, An23} and is called ``Schr\"odingerization'' or Linear Combination of Hamiltonian Simulations (LCHS).
The LCHS algorithm has an advantage over QLSAs in its higher success probability, which leads to a reduced number of calls to the initialization circuit.
Furthermore, the encoding of the LCHS algorithm is more straightforward compared to that of QLSAs.

We developed an efficient explicit implementation of the original LCHS algorithm\cite{An23} and simulated its numerical performance in Ref.~\onlinecite{Novikau24KvN}.
\ycb{However, we found that achieving modest \% levels of precision required a fairly large ancillary register for the construction of the kernel (corresponding to parameters $n_k\geq 7$ and $N_k\geq 128$ in the following).
The original LCHS algorithm was significantly improved by An, Childs, and Lin \onlinecite{An23impr} by exponentially reducing its query complexity with respect to the error, $\yerr$, to near-optimal scaling with precision, set out in Theorem~\ref{thm:near-optimal_LCHS} below.
Hence, our work here is motivated by the practical question of whether a more efficient LCHS implementation that has better dependence on precision can offer a significant savings in resources.}

\ycb{Hence, we test the new LCHS algorithm on the key point example of the linear advection-diffusion equation. This is a particularly important example of a dissipative differential equation because it is used to model physical processes in many different fields including astronomy, astrophysics, biology, chemistry, condensed matter physics, finance, fluid dynamics, and plasma physics.
In many contexts, the linearized version of advection-diffusion results from applying perturbation theory to a nonlinear problem, such as the Navier-Stokes equations or the Boltzmann kinetic equation. 
Yet, Liouville's theorem implies that the first-principles equations of motion have no nonlinearity for conservation of probability at either the classical or quantum mechanical level.
}

\ycb{
Hence, the linear advection-diffusion equation is also important for describing the first-principles probabilistic formulation of classical and semiclassical dynamics.~\cite{Joseph23JPA} 
This approach can be utilized to develop efficient quantum algorithms for evolving the probability distribution function for nonlinear dynamics based on the unitary Koopman-von Neumann (KvN) and Koopman-van Hove equations.~\cite{Joseph20,Joseph23,Novikau24KvN}
In fact, the main goal of \onlinecite{Novikau24KvN} was to develop an explicit high-performance quantum algorithm for simulating the KvN equation.
There, the approach relied on numerical dissipation, by adding either an upwind advection operator or a numerical diffusion operator to the generator, to eliminate numerical artifacts caused by the way in which perfectly unitary dynamics responds to sources and sinks.
Thus, simulating the evolution of the probability distribution of nonlinear dynamics is a potentially important future application of the efficient LCHS methods that we derive here.
}

\subsection{Main results}
The main focus of our work is to provide an efficient quantum circuit encoding of the LCHS algorithm and to test its performance in practice for an important point-example: the advection-diffusion equation.
Although the \ycb{seminal works~\onlinecite{An23, An23impr, Low25, Jin25}} offer an extensive analytical description of the LCHS method, they do not provide a systematic description of the LCHS circuit.
Our recent work~\onlinecite{Novikau24KvN} clearly demonstrated that the same algorithm described analytically can be encoded in a variety of ways that can significantly affect not only the total number of gates in the circuit but also the overall complexity of the encoded algorithm and its dependence on various parameters of the method under consideration.

In particular, the authors of Refs.~\onlinecite{An23, Sato24, Hu24Sch, Novikau24KvN} proposed constructing the core component of the LCHS/Schr\"odingerization algorithm as a sequence of controlled time-evolution operators. 
In contrast, in this work we reduce that component to a single Hamiltonian evolution implemented via a single QSP circuit.
The proposed encoding significantly simplifies the LCHS circuit compared to 
previous works.
More precisely, we demonstrate how a single block-encoding oracle can simultaneously represent an exponential number of matrices, enabling an exponential number of quantum Hamiltonian simulations (QHS) to be performed in parallel using a single QSP circuit.

The encoding used here builds on the coordinate transformation proposed in our previous work.\cite{Novikau24KvN} 
Here, we report on our discovery of how to leverage this transformation by incorporating it into the qubitization operator of the QSP circuit. 
This represents a significant improvement over the previous approach by eliminating the need for trotterization and significantly simplifying the selector circuit.
We also analyze the complexity of the LCHS algorithm based on our encoding technique and show that, unlike the analysis in Refs.~\onlinecite{An23, Novikau24KvN}, it eliminates the dependence of algorithmic accuracy on the trotterization step and matrix commutators.
Most importantly, the proposed encoding improves the scaling with respect to the simulated time.

We simulate the LCHS circuit on a digital emulator~\cite{Jones19, QuCF, code-OPT-LCHS} of fault-tolerant quantum computers and verify the circuit by modeling the advection-diffusion equation.
In these simulations, we demonstrate the near-optimal scaling of the improved LCHS algorithm with precision, its high success probability, and how its accuracy depends on key parameters such as the number of points in Fourier space and the length of the simulated time interval for a key numerical point example.
\ycb{In the interest of open and reproducible science,} all source files used in the simulations are available in Ref.~\onlinecite{code-OPT-LCHS} to allow the interested reader to reproduce our results.

\ycb{
Thus, in this work, we significantly improve on LCHS implementations in several important ways: 
\begin{enumerate}
\item By using the recently proposed class of kernels in \onlinecite{An23impr}, we obtain near-optimal convergence with precision.
\item We use a simple sinusoidal coordinate transformation to easily block-encode both the Hermitian and non-Hermitian parts of the operator and eliminate the need for trotterization in \onlinecite{Novikau24KvN}.
\item Thanks to this simplification, we can perform all necessary Hamiltonian simulations with a single quantum signal processing\cite{Low17} (QSP) circuit, requiring only two ancillae (without including block-encoding). 
\item Our integration method is equivalent to Fej\'er-Clenshaw-Curtis (FCC) quadrature, and Lemma~\ref{thm:FCC=CGL} shows that, for periodic and analytic functions, FCC is equivalent to Chebyshev-Gauss-Lobatto (CGL) quadrature at the Chebyshev extrema.
Thus, it has excellent convergence properties, especially in the exponentially convergent regime where it is optimally employed.
\item While the complexity analysis has recently been improved by focusing on the global convergence afforded by analytic function approximations~\cite{Low25, Jin25}, Lemma~\ref{thm:multiple-asymptotics} proves that this is generically only an intermediate asymptotic for kernels of interest.
Hence, regimes with power law scaling and nonlinear aliasing of the two factors in the LCHS integral require more careful consideration.  
\item We perform several additional optimizations to reduce the overall number of ancillae.
\item Theorem~\ref{thm:FCC-LCHS} proves that the FCC quadrature method has better error convergence, and, hence, better overall complexity than other LCHS circuit implementations proposed in the literature~\onlinecite{An23, An23impr, Low25}.
\end{enumerate}
Taken together, this series of improvements represents a significant advance over alternate suggested implementations of the LCHS circuit to date.
}

\ycb{
Achieving this wide range of improvements for a practical numerical example is important step forward for LCHS methods. 
Because our goal is to simulate the numerical performance, we have to focus on reducing the number of ancillary qubits as much as reasonably possible so that the problem instance fits within the memory of the classical computer that emulates the quantum circuit.
Moreover, this work and our previous work \onlinecite{Novikau24KvN} are the only in-depth numerical investigations of the numerical performance of the algorithm for the key point example of the advection-diffusion equation.
Because it is often difficult to perform a complete cost analysis for a complex problem, our work demonstrates the importance of performing careful numerical calculations that involve all key steps in the analysis.
}

\ycb{While the original works \onlinecite{An23,An23impr} suggested using both QSP and quantum singular value transformation~\cite{Gilyen19} (QSVT) for different aspects of an implementation of LCHS, they did not present an explicit block-encoding of the entire circuit, including the most important part, the selector.
In Ref.~\onlinecite{An23impr} (Section 4.2.1), a tentative description of an explicit circuit construction was proposed, but it is much more complicated than our version since it is based on composite Gaussian quadrature and their circuit includes several extra ancillary registers.
Moreover, ~\onlinecite{An23impr} explains that the most crucial step in LCHS is the implementation of the selector which is completely and definitively covered in our work.
}

\ycb{
Quite recently, after our work was completed and submitted, new approximate LCHS kernels that achieve the optimal scaling with precision for a linear differential equation solver\cite{Berry17}, $\oO(\log( \yerr^{-1}))$, 
were derived in Refs.~\onlinecite{Low25} and \onlinecite{Jin25}.
The key idea, originally explored in \onlinecite{Silva2022fourier} for the purpose of approximating the exponential function applied to an operator, is to use an ``analytic extension'' of the kernel function, based on analytic approximations to the step function.
While these newer works have made important contributions to understanding and improving the precision of approximate LCHS, only Ref.~\onlinecite{Low25} gives an explicit block-encoding for the selector. (Ref.~\onlinecite{Jin25} cites ~\onlinecite{An23impr} for the explicit circuit.)
The ``multiplexed block-encoding'' algorithm developed by Ref.~\onlinecite{Low25} Lemma 18 is more complicated than the construction proposed here, and requires several subroutines and additional ancillary registers, in fact, perhaps  doubling or even quadrupling the size of the register required for the kernel itself.
Thus, the key idea of using the sinusoidal coordinate transformation to perform FCC quadrature in combination with QSP for QHS is a simple and elegant solution that leads to a valuable reduction in both classical and quantum complexity that is important for practical implementations.
}

This paper is organized as follows.
In Sec.~\ref{sec:LCHS-basics}, we review the LCHS method with \ycr{the} 
improved \ycr{kernels} and explain in detail how this algorithm can be mapped on\ycr{to} a quantum circuit.
\ycb{Then, in Sec.~\ref{sec:complexity}, we consider the convergence of errors and how this affects the complexity of the LCHS algorithm. }
\ycr{Next, in} Sec.~\ref{sec:results}, we test the LCHS circuit by simulating
the advection-diffusion equation and analyze the algorithm scaling, precision, and the circuit success probability.
\ycr{Finally, in} Sec.~\ref{sec:conclusions}, we summarize the main results of this paper.
\section{
Efficient Implementation of Linear Combination of Hamiltonian Simulations (LCHS)
}\label{sec:LCHS}

\subsection{\ycb{Exact} LCHS 
}\label{sec:LCHS-basics}

\begin{figure}[!t]
\centering
\includegraphics[width=0.49\textwidth]{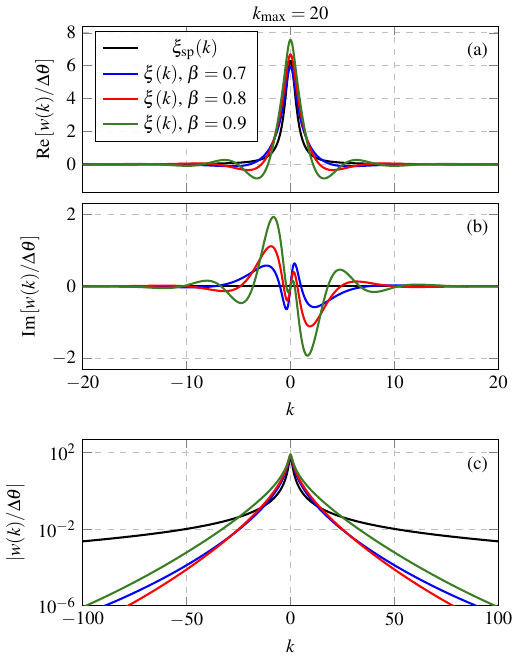}
\caption{
    \label{fig:LCHS-weights}
    Plots showing the real (a) and imaginary (b) components of the LCHS weights for the special case of the Cauchy kernel~\eqref{eq:xi-special} (black line) and the improved kernel~\eqref{eq:xi-general} with various $\beta$ (colored lines). (c): Absolute values of the LCHS weights.
}
\end{figure}
We consider a linear differential equation with a non-Hermitian time-independent generator $A$
\begin{equation}\label{eq:initial-diff-equ}
    \partial_t \psi(t,x) = - A \psi(t,x).
\end{equation}
Here, the evolution in time of the variable $\psi(t,x)$ is described by the nonunitary operator $e^{-At}$: 
\begin{equation}\label{eq:initial-value-problem}
    \psi(t,x) = e^{-At} \psi(0,x),
\end{equation}
where $\psi(0,x)$ represents the initial conditions
\ycb{and $\psi(t,x)$ is either real or complex. 
If the problem represents a partial differential equation (PDE), we assume that it has been discretized with $N_x$ grid points, and, hence, that the solution vector can be stored in quantum memory within a register of size $n_x=\log_2(N_x)$.
We note that both LCHS and Schr\"odingerization techniques have been developed for cases where the generator is time-dependent and where there is an inhomogeneous forcing in time; i.e. an equation of the form $d\psi/dt=-A(t)\psi(t,x)+b(t)$, but, in this work, we focus on the time-independent homogeneous case alone.
}

The initial-value problem~\eqref{eq:initial-value-problem} can be solved by exact LCHS\cite{An23, An23impr} which represents the nonunitary operator $e^{-At}$ as a weighted superposition of Hamiltonian evolutions integrated over an additional Fourier space
\begin{align}\label{eq:LCHS-theorem1}
    e^{- A t} = \int_\mathbb{R} \frac{\xi(k)}{1 - \yi k} e^{-\yi (A_H + k A_L) t}\diff k,
\end{align}
where the original generator $A$ has been separated into the Hermitian matrices
\begin{subequations}\label{eq:A-decomposition}
\begin{eqnarray}
    &&A = A_L + \yi A_H,\\
    &&A_L = (A + A^\dagger)/2,\label{eq:Ah}\\
    &&A_H = (A - A^\dagger)/ (2\yi) \label{eq:Aa}.
\end{eqnarray}
\end{subequations}
Equation~\eqref{eq:LCHS-theorem1} is valid as long as $A_L\ycb{\succeq 0}$ is positive semi-definite.
\ycb{This can be extended to non-positive matrices, $A_L$, by applying the LCHS method to $A'=A+|\lambda| I$, where $I$ is the identity matrix and $\lambda<0$ is a lower bound for the smallest, i.e. largest in magnitude but negative, eigenvalue of $A_L$.
To find the actual behavior in time, one must then rescale the results in time by multiplying by $e^{|\lambda| t}$ as a classical post-processing step.}

\ycb{An, Childs, and Lin~\onlinecite{An23impr} Result 1 (Theorem 6)  proved that many choices of kernel are possible as long as they satisfy a few conditions: (i)  decay $\abs{k}^a\abs{\xi} \leq C$ for some $a,C>0$, (ii) normalization  $\int \xi dk/(1-ik)=1$, (iii) continuity on the real line, and, (iv) analyticity in the lower half plane which allows one to close the contour integral in the lower half plane for $A_Lt \succeq  0$.
However, their no-go result \onlinecite{An23impr} Proposition 8 proves that, for kernels that satisfy these conditions and exactly satisfy Eq.~(3), it is not possible to achieve the optimal scaling with precision, $\oO(\log(\yerr^{-1}))$, expected for an ODE solver~\cite{Berry17}.
Yet, they showed that the family of kernels: }
\begin{equation}\label{eq:xi-general}
    \xi(k) = \frac{1}{2\pi e^{-2^\beta} \exp\left([1+\yi k]^\beta\right)},
\end{equation}
with the real scalar $\beta \in (0,1)$ \ycb{lead to near-optimal dependence on precision in the form $\oO(\log^{1/\beta}(\yerr^{-1}))$.
}

\ycb{They also proved~\cite{An23impr}  that the exact LCHS method based on kernel~\eqref{eq:xi-general} leads to a quantum linear differential equation solver that is nearly optimal in all respects.
For clarity, we now recall their Result 3 (Corollary 17) for a homogeneous differential equation with time-independent generator, $A$, which follows from their Result 2 (Theorem 15 and Corollary 16), which treats the inhomogeneous and time-dependent case.
}

\begin{theorem}[{\bf Near-Optimal LCHS Algorithm} \onlinecite{An23impr} Corollary 17] \label{thm:near-optimal_LCHS}
{ There is a quantum LCHS algorithm based on kernel~\eqref{eq:xi-general} that prepares the normalized solution of Eq.~\eqref{eq:initial-diff-equ} with $\Omega(1)$ success probability and a flag indicating success that uses $Q_A$ queries to the oracle encoding $A$ and $Q_{in}$ queries to the initial state preparation oracle, where}
\begin{align}
    Q_A&= \oO\left( \frac{\norm{\psi(0)}}{\norm{\psi(t)}} \|At\| \log^{1/\beta}(\yerr^{-1})\right)
    \\
    Q_{in} &= \oO\left( \frac{\norm{\psi(0)}}{\norm{\psi(t)}}\right).
\end{align}
\end{theorem}
\begin{proof}
The LCHS integral can be approximated optimally using LCU~\cite{Childs12} where the weight functions are computed optimally using QSVT\cite{Low19} and the success probability is boosted with amplitude amplification~\cite{Brassard02}. 
Given an efficient block-encoding of $A_H$ and $A_L$, each Hamiltonian simulation step can be performed optimally using either QSP~\cite{Low17} and qubitization~\cite{Low19} or the truncated Dyson series method~\cite{Low19Interaction}.
\end{proof}

\ycr{
In contrast, the original LCHS method described in Ref.~\onlinecite{An23}, which is equivalent to the mathematical formulation of the Schr\"odingerization algorithm in \onlinecite{Jin24}} employs the special case of the simple but suboptimal Cauchy kernel
\begin{equation}\label{eq:xi-special}
    \xi_{\rm sp}(k) = \frac{1}{\pi (1 + \yi k)}.
\end{equation}
Despite the simplicity, truncating the original LCHS integral at maximum values of $\pm \kmax$ yields a truncation error of order $\yerr\sim \oO(\kmax^{-1})$, so the cost to achieve a given precision scales as $\oO( \yerr^{-1})$. 
We tested the performance of this method in Ref.~\onlinecite{Novikau24KvN} for both advection-diffusion and upwind advection.
While we validated the algorithm and our implementation, we found that, in practice, significant resources were required  to reach a relative precision of order $10^{-2}$.
Thus, the main motivation for this work was to find an LCHS implementation with better performance for a practical numerical example.

\ycb{
One of our key contributions is the realization that to} efficiently map the LCHS equation~\eqref{eq:LCHS-theorem1} onto a quantum circuit, it is beneficial to apply the following coordinate transformation:\cite{Novikau24KvN}
\begin{eqnarray}\label{eq:coord-transf}
    k = \kmax \sin(\theta)
\end{eqnarray}
and to use this in the block-encoding of the Hamiltonian $C_k=kA_L+A_H$.
This choice allows one to recast the dependence on the Fourier coordinate $k$ as a trigonometric function, $\sin(\theta)$, which is  
much easier to block-encode \ycb{than $k$ itself, which technically requires implementing the $\arcsin(\theta)$ function.
}

\ycb{
Because $\arcsin(\theta)$ is not continuous, this implies that the Chebyshev coefficients only converge as $1/\Nq$, where, here, $\Nq$ is the number of terms in the series. 
Hence, to achieve a relative error of order $\yerr$, a large number of terms is required. 
While ~\onlinecite{McArdle2022quantum, Kane2025blockencodingbosons} have performed studies of the performance of this choice, and there may be methods for improvement using the ``number operator'' encoding of \onlinecite{Low25}, the scaling appears to be suboptimal. 
In contrast, the coordinate transformation of Eq.~\eqref{eq:coord-transf} is represented exactly by a single term.
Relative to our previous implementation \cite{Novikau24KvN}, this choice 
(1) eliminates the need for trotterization of $kA_L$ and $A_H$ and 
(2) significantly simplifies the implementation of the selector, $S$. 
(3) Finally, the choice of near-optimal kernel~\eqref{eq:xi-general} allows our new implementation to achieve near-optimal dependence on precision.
}

The LCHS computation~\eqref{eq:LCHS-theorem1} of the nonunitary operator $e^{- A t}$ is exact when one works in infinite Fourier space.
Yet, to simulate Eq.~\eqref{eq:initial-value-problem} numerically, one must truncate Fourier space, i.e. by imposing $|k| < \kmax$ and by discretizing the space with a grid with $N_k = 2^{n_k}$ points.
In order to approximate the integral, we discretize Fourier space through the transformation~\eqref{eq:coord-transf} via 
\begin{eqnarray}\label{eq:coord-transf-disc}
    k_j = \kmax \sin(\theta_j),
\end{eqnarray}
where the discretized angle $\theta_j$ is
\begin{equation}\label{eq:phi-grid}
    \theta_j = -\frac{\pi}{2} + j\Delta \theta,\quad \Delta \theta = \frac{\pi}{N_k-1}, \quad j = 0,1,\dots N_k-1. 
\end{equation} 
Hence, one obtains a discretized version of the initial-value problem~\eqref{eq:initial-value-problem}, 
\begin{equation}\label{eq:psi-LCHS}
    \psi(t,x) = U_{\rm LCHS} \psi(0,x) + \yerr,
\end{equation}
where $\yerr$ is the combination of discretization and truncation error.
The discretized LCHS operator $U_{\rm LCHS}$ approximates the nonunitary operator $e^{-At}$ as
\begin{equation}\label{eq:LCHS-discr}
    e^{- A t} \approx U_{\rm LCHS} = \sum_{j=0}^{N_k-1} w_j V_j(t),
\end{equation}
where each unitary $V_j$,
\begin{equation}\label{eq:Vj}
    V_j(t) = e^{-\yi C_j t},
\end{equation}
depends on the $\theta$-dependent Hermitian matrix $C_j$,
\begin{subequations}\label{eq:C}
\begin{eqnarray}
    &&C_j = A_H +  \sin(\theta_j)\, B_{\rm m},\\
    &&B_{\rm m} = \kmax A_L,\label{eq:Bm}
\end{eqnarray}
\end{subequations}
and the sum~\eqref{eq:LCHS-discr} is weighted by the complex coefficients $w_j$,
\begin{equation}\label{eq:wj}
    w_j = \frac{\kmax\cos(\theta_j)\,\Delta\theta\,\xi(k_j)}{1-\yi k_j}.
\end{equation}
These weights become real when the Cauchy kernel~\eqref{eq:xi-special} is applied.
The shape of the LCHS weights $w_j$ computed with the improved and Cauchy kernels, Eq.~\eqref{eq:xi-general} and Eq.~\eqref{eq:xi-special}, correspondingly, are shown in Fig.~\ref{fig:LCHS-weights}.
One can see that the real and imaginary components of the complex weights built with the improved kernel $\xi$ are functions of definite parity. 
This means that each component can be constructed by using a single QSVT circuit. 
Also, as can be seen from Fig.~\ref{fig:LCHS-weights}c, the improved kernel causes the weights to decay exponentially faster than in the special case~\eqref{eq:xi-special}.
According to Ref.~\onlinecite{An23impr} and as is also demonstrated later in this paper, this results in the exponential decay of the truncation error $\yerr$ with $\kmax$:
\begin{equation}\label{eq:err-scaling-impr-theory}
    \yerr = \oO\left(e^{-\ycb{\oO(\kmax^\beta})}\right).
\end{equation}
At the same time, the truncation error in the discretized LCHS computation~\eqref{eq:psi-LCHS} with the Cauchy kernel $\xi_{\rm sp}$ is only inversely proportional to $\kmax$:\cite{An23, Novikau24KvN}
\begin{equation}\label{eq:kmax-conv-sp}
    \yerr = \oO\left(\kmax^{-1}\right).
\end{equation}

\subsection{Comparison of Exact and Approximate LCHS}
\ycb{After this work was submitted, more general kernels that achieve more optimal scaling with precision 
were derived in Refs.~\onlinecite{Low25} and \onlinecite{Jin25}.
In fact, Ref.~\onlinecite{Silva2022fourier}, previously explored the same idea of using an ``analytic extension'' of a Fourier series for the goal of computing the exponential function of an operator  using the QSP-based algorithm they derived for constructing an arbitrary Fourier series of a block-encoded operator, and, studied the performance numerically.}

\ycb{
The key to avoiding the no-go theorem of \onlinecite{An23impr} is to search for kernels that approximate Eq.~\eqref{eq:LCHS-theorem1} to precision $\oO(\yerr)$.
The simplest versions of such kernels approximate exponential decay in time with a function that is holomorphic in the entire complex plane (an entire function); i.e. they either approximate the functions $e^{-|x|}$ or $e^{\mp x}\Theta(\pm x)$ where $\Theta(x)$ is the step function, by using the error function, ${\mathrm{erf}}\,(x)$, as an analytic approximation to the step function. 
While approximations of this type had already been explored in \onlinecite{Silva2022fourier} and while the error analysis of \onlinecite{Silva2022fourier} and \onlinecite{Low25} provide similar estimates, Refs.~\onlinecite{Low25,Jin25} made significant and valuable contributions in terms of analyzing the appropriate form of such approximations and the associated complexity in the context of approximate LCHS.
Interestingly enough, an extensive search for optimality in  \onlinecite{Low25} (see their Table~3) found that approximate LCHS kernels that were numerically optimized to have the best dependence on precision have the form of a rational function multiplied by the simple weight function and do not use the error function at all.
We refer the reader to Refs.~\onlinecite{Silva2022fourier,Low25,Jin25} for explicit derivations of examples of these kernels as well as proofs of their existence and optimality.
}

\ycb{
While it is well beyond the scope of this work to test the performance of these new kernels, our implementation is agnostic of the particular form of the choice of kernel and will work equally well for these improved kernel choices.
The main difference is that the dependence of the query complexity with precision goes from $\log^{1/\beta}(\yerr^{-1})$ to $\log(\yerr^{-1})$, where, for the example studied here and as first observed in \onlinecite{An23impr}, a near-optimal range is $\beta\in[0.7,0.8]$, i.e. $1/\beta \in [1.25, 1.43]$.
}

\ycb{A relatively straightforward explicit block-encoding for the selector, which requires a block-encoding of $C_k=kA_L+A_H$, was given in Ref.~\onlinecite{Low25}. 
Their approach directly encodes the number state $\sum_k k\left|k\right>$ and requires a second ancillary qubit register of size $n_k=\log_2(N_k)$, which doubles the size of the total ancillary memory register for LCHS to $\sim 2n_k$.
This procedure requires an additional $2\times \oO(n_k)$ two-qubit gates to apply a unitary that constructs the state $k^{1/2}\left|k\right>$ as well as the inverse, to ultimately apply the $k\left|k\right>\left<k\right|$ operator.
Then, one must perform a controlled SWAP of the two $k$ registers at an additional cost of $\oO(3n_k)$; however, for fixed hardware layout, the SWAP cost can be much higher.
As discussed more fully in Sec.~\ref{sec:further-considerations}, the total ancillary register required for this part can be as large as $4n_k$, depending on whether the extra required ancillary registers can be reused.
}

\ycb{
In contrast, the coordinate transformation of Eq.~\eqref{eq:coord-transf} only requires $\oO(1)$ ancillary qubits, and, as we shall see later, because it  represents a smooth transformation of variables, it only has an $\oO(1)$ impact on the cost of approximating the weight function using QSVT.
Finally, the cost of constructing the $\sin(\theta)$ function (see  Fig.~\ref{circ:sin}) is precisely $n_k$ controlled rotations and $2$ single-qubit rotations.
}

\subsection{\ycb{Explicit} LCHS Circuit}\label{sec:circuit}

\begin{figure}[!t]
\centering
\includegraphics[width=0.49\textwidth]{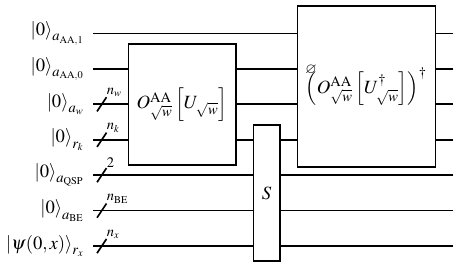}
\caption{
    \label{circ:LCHS}
    LCU circuit for solving the LCHS equation~\eqref{eq:psi-LCHS}.
    The selector is shown in Fig.~\ref{circ:selector}.
    \ycb{The $O^{AA}_{\sqrt{w}}$ operators perform amplitude amplification (AA) of the $U_{\sqrt{w}}$ operators which construct the weights via QSVT.
    These operators are explicitly defined in \onlinecite{Novikau24KvN} (Fig.~6 and 7, respectively).}
    The symbol $\varnothing$ indicates that the corresponding qubit is not used by the indicated subcircuit.
}
\end{figure}
\begin{figure}[!t]
\centering
\subfloat[LCHS selector as a QSP circuit]{\includegraphics[width=0.40\textwidth]{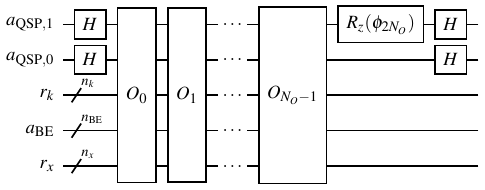}\label{circ:qsp}}\\
\vspace{0.2cm}
\subfloat[Operator $O_j$]{\includegraphics[width=0.49\textwidth]{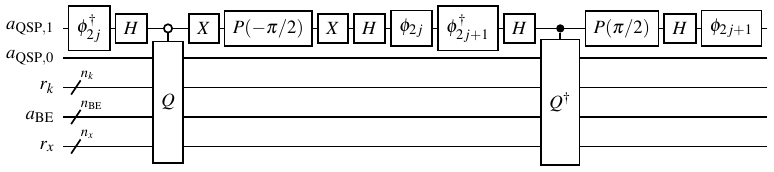}\label{circ:qsp-part}}\\
\caption{
    \label{circ:selector}
    (a) 
    The selector circuit $S$ for computing the $N_k$ unitaries $V_j$ [Eq.~\eqref{eq:Vj}].
    The selector is represented by a single QSP\cite{Low17, Low19} circuit consisting of $N_O$ operators $O_j$, where the number $N_O$ depends on the simulated time interval $t$.
    (b) Circuit for the operators $O_j$ for $j = 0,1,\dots N_O-1$.
    Each $O_j$ operator depends on two QSP angles, $\phi_{2j}$ and $\phi_{2j+1}$.
    Here, the gates schematically denoted as $\phi_l$ indicate the rotations $R_z(\phi_l)$.
    The iterate $Q$ is shown in Fig.~\ref{circ:qubitization}
}
\end{figure}
\begin{figure}[!t]
\centering
\subfloat[Iterate $Q$]{\includegraphics[width=0.40\textwidth]{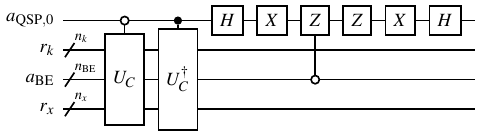}\label{circ:iterator}}\\
\vspace{0.2cm}
\subfloat[Block-encoding oracle $U_C$]{\includegraphics[width=0.36\textwidth]{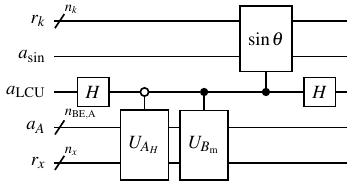}\label{circ:be-general}}\\
\vspace{0.2cm}
\subfloat[
$\sin\theta$ circuit]{\includegraphics[width=0.40\textwidth]{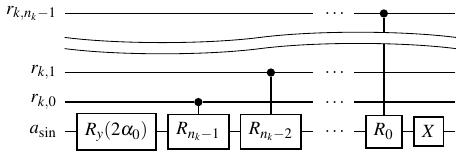}\label{circ:sin}}\\
\caption{
    (a) 
    The iterate circuit $Q$ used in the selector of Fig.~\ref{circ:qsp-part}.
    Note that the qubitization is performed only with respect to the ancillary register $a_{\rm BE}$ independently of the register $r_k$, although the latter is also used by the block-encoding oracle $U_C$.
    (b) 
    The block-encoding oracle $U_C$ encoding $C_j$ in Eq~\eqref{eq:C}.
    The implementation of the subcircuits $U_{A_H}$ and $U_{B_{\rm m}}$ depends on the structure of $A$.
    The registers $a_A$, $a_{\rm LCU}$, and $a_{\rm sin}$ are different parts of the register $a_{\rm BE}$ used in Fig.~\ref{circ:iterator}.
    (c) Circuit encoding the function $\sin(\theta_j)$ for $j = 0,1,\dots N_k - 1$.
    Here, $R_l = R_y(2\alpha_1/2^l)$, $\alpha_0 = -\pi/2$, and $\alpha_1 = |\alpha_0| N_k / (N_k-1)$. 
}
\label{circ:qubitization}
\end{figure}
\subsubsection{Overview}
\ycb{An entire quantum linear differential equation solver algorithm (QLDS) based on LCHS consists of amplitude amplification (AA) \cite{Brassard02} of the overall LCHS circuit and then a measurement of interest.
As will be explained further below, AA is essential for the QLDS to solve a dissipative differential equation because the success probability, which tracks the amplitude, tends to decay in time.
The measurement step is typically preceded by a series of state preparations required to measure the physical observable of interest, e.g. see ~\onlinecite{Joseph20} for an explicit discussion of how to measure an observable of interest, and typically also requires AA to boost the success probability and/or to accurately measure an amplitude.
In this work, we only focus on an efficient construction and emulation of the core LCHS circuit alone because the AA and measurement steps are standard.
Interestingly enough, for the numerical example over the specific time period studied below, we find that the overall success probability is actually $\sim$0.2, and, since LCHS terminates with flag indicating success, AA is not strictly required.
}

\subsubsection{\ycb{ LCU implementation of LCHS}}
\begin{figure}[!t]
\centering
\subfloat[Block-encoding of $A_H$]{\includegraphics[width=0.50\textwidth]{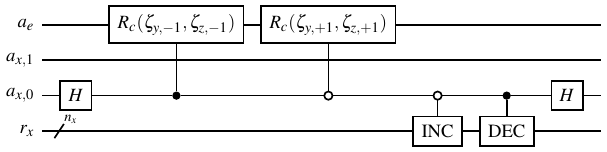}\label{circ:be-AH}}\\
\vspace{0.2cm}
\subfloat[Block-encoding of $A_L$]{\includegraphics[width=0.50\textwidth]{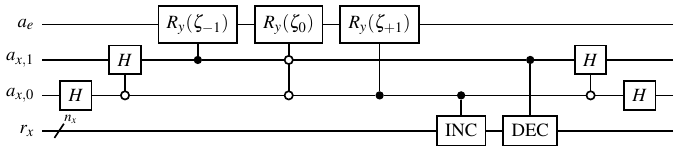}\label{circ:be-AL}}
\caption{
    \label{circ:be-AH-AL}
    (a) 
    The oracle encoding the matrix $A_H$, Eq.~\eqref{eq:A-decomposition}, for the advection-diffusion Eq.~\eqref{eq:ADE}. 
    (b) 
    The oracle encoding the matrix $A_L$. 
    The circuits for the incrementor (INC) and decrementor (DEC) can be found in Ref.~\onlinecite{Novikau22}, Figure 14; these circuits scale as $\oO(n_x)$.
    The gate schematically denoted by $R_c$ is described in Eq.~\eqref{eq:Rc}.
    The $\zeta_j$ parameters of the rotation operators are described in Eq.~\eqref{eq:zeta}.
}
\end{figure}

The discretized LCHS transformation is expressed in the form of a weighted sum, Eq.~\eqref{eq:LCHS-discr}.
As such, it can be mapped on a Linear Combination of Unitaries (LCU) circuit \ycb{of the standard form
\begin{align}
LCU= \yOwc \circ S \circ \yOw,
\end{align}
}
schematically shown in Fig.~\ref{circ:LCHS}.
There, the oracles $\yOw$ and $\yOwc$ compute the LCHS weights~\eqref{eq:wj}, and the selector oracle $S$, shown in Fig.~\ref{circ:selector}, computes the $N_k$ unitaries 
\ycb{
$V_j$, Eq.~\eqref{eq:Vj},
through a single QSP circuit, effectively in parallel.
The selector uses the operators, $O_j$, shown in Fig.~\ref{circ:qsp-part}, which contain the iterate $Q$ shown in Fig.~\ref{circ:iterator}.
The iterate depends on the block-encoding of $C_j$ given by $U_C$ and $U_C^\dagger$.
To encode $C_j$ efficiently, one needs access to oracles $U_{A_H}$ and $U_{B_m}$ that block-encode $A_H$ and $B_m=\kmax A_L$, as well as the $\sin(\theta)$ circuit.
All necessary subcircuits are explicitly shown in Fig.~\ref{circ:qubitization}.
The oracles $U_{A_H}$ and $U_{B_m}$ for our particular case of the linear advection-diffusion equation are given in Fig.~\ref{circ:be-AH-AL}.
The resources used by our LCHS circuit are summarized in Tables~\ref{tab:LCHS-subcircuits} and~\ref{tab:LCHS-circuit-size}.
}

The register $r_k$ is used to encode the dependence on the Fourier coordinate $k_j$, expressed as the sine function of the angles $\theta_j$ according to Eq.~\eqref{eq:coord-transf-disc}.
The ancillary qubits $a_{\rm QSP}$ and $a_{\rm BE}$ are used for the construction of the selector and for block-encoding the matrices $C_j$, Eq.~\eqref{eq:C}.
The ancillary register $a_w$ with $n_w$ qubits is used by both oracles $\yOw$ and $\yOwc$ for computing the LCHS weights.
In addition, the oracles $\yOw$ and $\yOwc$ use the ancillae $a_{{\rm AA}, 0}$ and $a_{{\rm AA}, 1}$, correspondingly, for AA of the computed weights.
Finally, the register $r_x$ encodes the initial condition $\psi(0,x)$, which must be precomputed by an additional initialization subcircuit.
The same register outputs the result, $\psi(t,x)$, entangled with the zero state of all ancillary qubits and the register $r_k$. 

\subsubsection{\ycb{Weights prepared by QSVT}}
The complex LCHS weights can be computed by two QSVT circuits combined by an LCU. 
There, each QSVT can be implemented in the manner explained in detail in Ref.~\onlinecite{Novikau24KvN} \ycb{and will require $n_w = 3$ qubits in register $a_w$, taking into account the LCU procedure necessary to combine complex weights from real functions returned by QSVT.}
In particular, the oracle $\yOw$ includes as a subcircuit the oracle $U_{\sqrt{w}}$, which computes the complex 
\ycb{state 
\begin{align}
U_{\sqrt{w}}\left|0\right>_{r_k}=\sum_j \sqrt{w_j}\left|j\right>_{r_k}
\end{align}
}
by using the combined QSVT-LCU circuit.
An alternative approach for the computation of $\sqrt{w_j}$ is a tensor-network-based technique described in Ref.~\onlinecite{Sato24}.
In our numerical emulations, we use a direct exact brute-force computation of the LCHS weights to minimize the number of ancillary qubits in the circuit. 
\ycb{In this case, only $n_w = 1$ ancilla is needed in register $a_w$, and} 
the oracle $U_{\sqrt{w}}$ is constructed as a sequence of $N_k$ combined rotations $R_y(\phi_{y,j})R_z(\phi_{z,j})$ where $\phi_{y,j} = 2 \arccos(|\sqrt{w_j}|)$ and $\phi_{z,j} = - 2 \arg{\left(\sqrt{w_j}\right)}$.

After the weight computation, one should apply AA to amplify the success probability of the oracle $U_{\sqrt{w}}$, thereby boosting the success probability of the entire LCHS circuit~\ref{circ:LCHS}.
The amplification is implemented by the oracle $\yOw$ and can be accomplished by using the standard AA technique.\cite{Brassard02}
\ycb{The need for AA in these steps} was explained in detail in Ref.~\onlinecite{Novikau24KvN}.
In particular, AA requires $2 N_{\rm AA}$ repetitions of the subcircuit $U_{\sqrt{w}}$.
For instance, for $\kmax = 40$, one has $N_{\rm AA} = 30$ and $N_{\rm AA} = 43$ for $n_k = 11$ and $n_k = 12$, correspondingly.
An important remark here is that AA does not concern the selector $S$ which is the most computationally intensive component of the LCHS circuit.
Therefore, one performs only $\oO(N_{\rm AA})$ repetitions of $U_{\sqrt{w}}$ without touching the selector.
Apart from this, since $S$ makes the dominant contribution to the complexity of the LCHS circuit and scales with time, the choice in the implementation of $U_{\sqrt{w}}$ has a small effect on the length of the entire LCHS circuit.

\subsubsection{\ycb{Selector implementation with a single QSP circuit}}
The selector $S$ is implemented as a single QSP\cite{Low17, Low19} circuit shown in Fig.~\ref{circ:selector} which is only possible due to the coordinate transformation~\eqref{eq:coord-transf}.
In particular, the QSP circuit computing the unitaries $V_j$ depends on the block-encoding oracle $U_C$ shown in Fig.~\ref{circ:be-general} encoding the Hermitian $C_j$, Eq.~\eqref{eq:C}, where $C_j$ is normalized as 
\begin{subequations}\label{eq:Cmax-norm}
\begin{eqnarray}
    &&C_j \to C_j/\|C_{\rm max}\|,\\
    &&\|C_{\rm max}\| = \|A_H + \kmax A_L\|.
\end{eqnarray}
\end{subequations}
Each matrix $C_j$ is represented by a sum of two Hermitians, $A_H$ and $B_{\rm m}$, and depends on the sine function $\sin(\theta_j)$.
Therefore, the oracle $U_C$ is constructed as an LCU combining two subcircuits, $U_{A_H}$ and $U_{B_{\rm m}}$, and an extra subcircuit computing the sine function, Fig.~\ref{circ:sin}.
In particular, the coordinate transformation~\eqref{eq:coord-transf} is necessary to significantly simplify the encoding of the dependence on the Fourier coordinate $k$ by using the subcircuit~\ref{circ:sin}.
The implementation of the block-encoding oracles $U_{A_H}$ and $U_{B_{\rm m}}$ is problem-specific and depends on the structure of the original non-Hermitian generator $A$ used in Eq.~\eqref{eq:initial-diff-equ}.

\ycb{
Our proposed implementation of the selector is much more compact than that given in previous works such as Refs.~\onlinecite{Hu24Sch, Sato24, Novikau24KvN}.
In particular, instead of using trotterization where each time step is simulated by a sequence of $n_k$ controlled Hamiltonian simulations, our encoding performs a single Hamiltonian evolution. 
This is achieved by hiding the dependence on the Fourier coordinate $k$ within the iterate, $Q$.
As shown in Sec.~\ref{sec:selector-scaling}, this eliminates the dependence of the overall accuracy on the trotterization step and the commutator of the matrices $A_L$ and $A_H$.
Thus, the selector achieves linear scaling with time.
}

\subsubsection{\ycb{Block-Encoding of Hamiltonian $C=A_H+kA_L$ }}
\ycb{Efficient block-encodings are necessary for quantum advantage and block-encodings for many important PDEs have been derived.\cite{Novikau24KvN,Guseynov2025block}
The block-encoding oracles $U_{A_H}$ and $U_{B_m}$ can be constructed following the procedure thoroughly described in Refs.~\onlinecite{Novikau23, Novikau24-EVM, Novikau24KvN}.
Our definition of block-encoding is standard and is, for example, given in in the seminal work Ref.~\onlinecite{Gilyen19} Def.~24.
Let us assume that we need to encode a matrix $M$.
Then, the goal of the block-encoding is to construct a circuit represented by the unitary $U_M$ such that 
\begin{equation}\label{eq:UM}
U_M = 
    \begin{pmatrix}
        \frac{M}{\|M\|\varsigma} & \cdot \\
        \cdot & \cdot
    \end{pmatrix},
\end{equation}
where $\|M\|$ is the matrix spectral norm, the scalar $\varsigma$ is determined by the matrix sparsity $\varsigma^\prime$, which is the maximum number of nonzero elements in any row or column of the matrices $A_L$ and $A_H$.
The upper bound of $\varsigma$ can be estimated as
\begin{equation}
    \varsigma = 2^{\yceil{\log_2\varsigma^\prime}}.
\end{equation}
The oracle $U_M$ acts on the registers $a_A$ and $r_x$:
\begin{equation}
    \left|\frac{M_{rc}}{\|M\|\varsigma} - \bra{c}_{r_x}\bra{0}_{a_A} U_M \ket{r}_{r_x}\ket{0}_{a_A} \right| \leq \epsilon_{\rm BE},
\end{equation}
where $r$ and $c$ are the row and column indices, respectively, of a nonzero element of the matrix $M$, and $\epsilon_{\rm BE}$ is the block-encoding error.
}

\subsubsection{\ycb{Qubitization}}
Qubitization is used to transform each eigenvalue of the block-encoded matrix $C_j$ within its own Hilbert subspace.
Thus, all matrix eigenvalues are transformed in their own disjoint two-dimensional subspaces which allows us to build different powers of the encoded matrix.
In this way, we obtain a Grover-like search parallelized over all matrix eigenvalues.
\ycb{It is important to emphasize that, in this paper,  we employ fully coherent} QSP in the formulation described in Ref.~\onlinecite{Low19}.
\ycb{Thus, there is no impact on the success probability and this} step does not require AA.

The oracle $U_C$ is a part of the iterate $Q$ shown in Fig.~\ref{circ:iterator}, which performs qubitization.\cite{Low19}
\ycb{This oracle can be constructed as
\begin{equation}
    Q = (U_R\otimes I_{r_x, r_k}) U_C^\prime,
\end{equation}
where $U_R$ is the reflector operator
\begin{equation}\label{eq:reflector}
    U_R = 2\ket{+}_{a_{{\rm QSP},0}}\ket{0}_{a_{\rm BE}}\bra{0}_{a_{\rm BE}}\bra{+}_{a_{{\rm QSP},0}} - I_{a_{{\rm QSP},0},a_{\rm BE}},
\end{equation}
and $U_C^\prime$ is a combination of two controlled block-encoding oracles $U_C$
\begin{equation}\label{eq:u-prime}
    U_C^\prime = \ket{0}_{a_{{\rm QSP},0}}\bra{0}_{a_{{\rm QSP},0}}\otimes U_C + \ket{1}_{a_{{\rm QSP},0}}\bra{1}_{a_{{\rm QSP},0}}\otimes U_C^\dagger.
\end{equation}
Also, $I_{r_x, r_k}$ is the unit operator that acts on the registers $r_x$ and $r_k$, and $I_{a_{{\rm QSP},0},a_{\rm BE}}$ is the unit operator that acts on the registers $a_{{\rm QSP},0}$ and $a_{\rm BE}$.
}

More precisely, qubitization is implemented by using the reflection operator $U_R$, represented by the last six gates in Fig.~\ref{circ:iterator}.
The latter must perform the reflection with respect to any ancillary qubit state orthogonal to the state entangled with the encoded matrices $C_j$.
In our case, each matrix $C_j$ for a particular $j$ is entangled with the state $\ket{j}_{r_k}\ket{0}_{a_{\rm BE}}$.
Since we want to implement the qubitization for the matrices $C_j$ with all $j = 0, 1, \dots N_k-1$ in parallel, we perform the reflection only with respect to $\ket{0}_{a_{\rm BE}}$ independently of the state of the register $r_k$.
This ensures that the QSP circuit computes the unitaries $V_j$ for all $j$ at once.
In other words, the coordinate transformation~\eqref{eq:coord-transf} and the oracle~\ref{circ:qubitization} allow us to perform $N_k$ Hamiltonian simulations in parallel by using a single QSP circuit.


\begin{table*}
\begin{center}
    \begin{tabular}{|c|c|c|c|}
       \hline
       \bf Figure & \bf Circuit  & \bf Purpose & \bf Complexity Scaling \\
       \hline 
        \ref{circ:LCHS}        & LCHS           & LCHS implemented as an LCU circuit, Eq.~\eqref{eq:LCHS-discr} 
            & $\oO\left(\frac{\|\psi(0)\|}{\|\psi(t)\|}\left(Q_{\rm sel} + Q_w\right)\right)$, Eq.~\eqref{eq:Q_LCHS} \\
        \ref{circ:qsp}         & $S$            & selector implemented as a QSP\cite{Low19} circuit 
            & $Q_{\rm sel} = \oO\left(Q_{\rm BE}\left[\kmax\ybCm t + Q_{\varepsilon,\rm QSP}\right]\right)$, Eq.~\eqref{eq:sel-scaling-final} \\
        \ref{circ:qsp-part}    & $O_j$          & a step in the QSP procedure 
            & $\oO(Q_{\rm BE})$ \\
        \ref{circ:iterator}    & $Q$            & qubitization iterate\cite{Low19}
            & $\oO(Q_{\rm BE})$ \\
        \ref{circ:be-general}  & $U_C$          & encode $C_j$, Eq.~\eqref{eq:C} 
            & $Q_{\rm BE} = \oO(n_k + {\rm poly}(n_x, \varsigma^\prime))$, Eq.~\eqref{eq:Q-be}\\
        \ref{circ:sin}         & $\sin(\theta)$ & compute $\sin(\theta_j)$ 
            & $\oO(n_k)$\\
        \ref{circ:be-AH} & $U_{A_H}$      & encode $A_H$, Eq.~\eqref{eq:A-decomposition} 
            & $\oO({\rm poly}(n_x, \varsigma^\prime))$ \\
        \ref{circ:be-AL} & $U_{B_m}$      & encode $B_m$, Eq.~\eqref{eq:Bm}
            & $\oO({\rm poly}(n_x, \varsigma^\prime))$ \\
        \hline 
        Fig. 6 in Ref.~\onlinecite{Novikau24KvN} & $O^{\rm AA}_{\sqrt{w}}$ in circuit~\ref{circ:LCHS} & amplitude amplify weights
            & $Q_w = \oO\left(n_k\left[\kmax^{3/2}m_\varepsilon\yerr^{-1}\right]\right)$, Eq.~\eqref{eq:w-scaling} \\
        Fig. 7 in Ref.~\onlinecite{Novikau24KvN} & $U_{\sqrt{w}}$ in circuit~\ref{circ:LCHS} & encode weights $w_j$
            & $\oO(n_k \kmax m_\varepsilon \yerr^{-1} )$, Eq. 53 in Ref.~\onlinecite{Novikau24KvN}\\
        \hline
    \end{tabular}
\end{center}
\caption{
    \ycb{LCHS subcircuits, their purpose, and scaling. The numerical implementation of the entire LCHS circuit can be found in Ref.~\onlinecite{code-OPT-LCHS}.
    The implementation of the block-encoding oracles $U_{A_H}$ and $U_{B_m}$ (Fig.~\ref{circ:be-AH-AL}) is specific to the linear uniform advection-diffusion problem and will differ for other problems. 
    The structure of the rest of the LCHS circuit (Figs.~\ref{circ:LCHS},~\ref{circ:selector}, and~\ref{circ:qubitization}) is problem-independent.}
}
\label{tab:LCHS-subcircuits}
\end{table*}

\begin{table*}
\begin{center}
    \begin{tabular}{|c|c|c|l|l|}
    \hline
    \bf Circuits & \bf Register  & \bf Qubits & \bf Use & \bf Purpose \\
\hline
    \ref{circ:LCHS}-\ref{circ:be-AH-AL} 
        & $r_x$       
        & $n_x$     
        & I/O 
        & input $\psi(0,x)$ and output $\psi(t,x)$ entangled with the zero state of all ancillae \\
    \ref{circ:LCHS}                     
        & $a_w$  
        & $n_w = 3$ 
        & ancilla 
        & LCU-QSVT computing weights $w_j$\\
    \ref{circ:LCHS}                     
        & $a_{\rm AA}$       
        & $n_{{\rm AA}, w} = 2$         
        & ancilla 
        & amplitude amplification of QSVT \\
    \hline
    \ref{circ:LCHS} - \ref{circ:qubitization} 
        & $r_k$         
        & $n_k$ 
        & ancilla 
        & encode the integer $j$, Eq.~\eqref{eq:phi-grid}, to address angle $\theta_j$ \\
    \ref{circ:LCHS} - \ref{circ:iterator}     
        & $a_{\rm QSP}$ 
        & $2$   
        & ancilla 
        & construct the QSP and the iterate circuits for the implementation of selector $S$  \\
    \ref{circ:LCHS} - \ref{circ:iterator}     
        & $a_{\rm BE} = a_{\rm sin} + a_{\rm LCU} + a_{A}$ 
        & $n_{\rm BE}=2 + n_{\rm BE, A}$ 
        & ancilla 
        & block-encode $C_j$ \\
    \hline
    \ref{circ:be-general} and \ref{circ:sin} 
        & $a_{\rm sin}$ 
        & 1                     
        & in ancilla $a_{\rm BE}$ 
        & for computing $\sin(\theta_j)$ \\
    \ref{circ:be-general}
        & $a_{\rm LCU}$ & 1                     
        & in ancilla $a_{\rm BE}$ 
        & for computing LCU, i.e. the sum operation in Eq.~\eqref{eq:C} \\
    \ref{circ:be-general}
        & $a_{A} = a_e + a_x$ 
        & $n_{\rm BE, A}$ 
        & in ancilla $a_{\rm BE}$ 
        & block-encoding $B_m$ and $A_H$\\
    \hline
    \ref{circ:be-AH-AL} 
        & $a_e$ 
        & 1 
        & in ancilla $a_{A}$ 
        & for computing nonzero elements in the matrices $A_H$ and $B_m$\\ 
    \ref{circ:be-AH-AL} 
        & $a_x$ 
        & 2 
        & in ancilla $a_{A}$
        & encoding column indices of nonzero elements in the matrices $A_H$ and $B_m$ \\
    \hline
    \end{tabular}
\end{center}
\caption{
    \ycb{Listing of the size and purpose of LCHS state and ancillary registers.
    Aside from the register $r_k$, $9 + n_{\rm BE, A}$ ancillae are required for this implementation of LCHS.
    In the case of the linear uniform advection-diffusion equation, $n_{\rm BE, A} = 3$.
    This does not account for additional registers that may be required for pre-processing steps such as the initialization of the circuit and the computation of initial conditions or for post-processing steps such as amplitude estimation and measurement of physical observables.}
}
\label{tab:LCHS-circuit-size}
\end{table*}


\section{LCHS Convergence \& Complexity \label{sec:complexity}}

\subsection{ Overall Complexity}\label{sec:complexity-scaling}
\subsubsection{ Error Components \label{sec:error-components}}
\ycb{
The total error for the LCHS algorithm has multiple components, $\yerr=\varepsilon_{\rm cont}+\varepsilon_{\rm disc}$.
First, an approximate LCHS method has the error, $\varepsilon_{\rm cont}$, because the choice of continuous kernel only approximates exact LCHS. 
Second, the error associated with discretizing the integral, $\varepsilon_{\rm disc}=\varepsilon_{\rm trunc}+\varepsilon_{\rm quad}$, has two parts, the truncation error, $\varepsilon_{\rm trunc}$, due to cutting off the integral at $\pm\kmax$, and the quadrature error, $\varepsilon_{\rm quad}$, due to discretizing the integral with a finite set of points.
Finally, another source of error results from the subroutine that computes the integrand, $\varepsilon_{\rm func}=\varepsilon_{\rm HS}+\varepsilon_{w}$, which, in turn, is the sum of errors in Hamiltonian simulation, $\varepsilon_{\rm HS}$, and the computation of the  weight function, $\varepsilon_w$.
Thus, the total error can be summarized as
\begin{align}\label{eq-err-lchs-decomp}
\yerr&=\varepsilon_{\rm cont}+\varepsilon_{\rm disc}+\varepsilon_{\rm func}\nonumber\\
&=\varepsilon_{\rm cont}+\varepsilon_{\rm trunc}+\varepsilon_{\rm quad}+\varepsilon_{\rm HS}+ \varepsilon_w.
\end{align}
Clearly, one can achieve a given overall LCHS error by requiring each component  to satisfy $\varepsilon_*\leq \yerr/m_\varepsilon$, where $m_\varepsilon$ is the number of terms in the total error, e.g.  $m_\varepsilon=5$ for the expression above.
Finally, for the example studied here, we use an exact near-optimal LCHS method, so $\varepsilon_{\rm cont}=0$ and, thus, $m_\varepsilon=4$ in the case of exact LCHS.
}

\subsubsection{ Complexity of the Entire LCHS Circuit}
\label{sec:LCHS-scaling}
 
\ycb{ The complexity of the entire LCHS circuit is set by the adding the cost of the selector, $Q_{\rm  sel}$ to the cost of computing the weights, $Q_w$, via QSVT, and the cost needed to amplitude amplify the success probability.
Thus, } the scaling of the entire LCHS algorithm is
\begin{equation}\label{eq:Q_LCHS}
    Q_{\rm LCHS} = \oO\left(\frac{\|\psi(0)\|}{\|\psi(t)\|}\left(Q_{\rm sel} + Q_w\right)\right),
\end{equation}
where the term $Q_{\rm sel}$, Eq.~\eqref{eq:sel-scaling-final}, is the dominant one because of its dependence on the simulated time $t$.
Here, the multiplicative factor $\|\psi(0)\|/\|\psi(t)\|$ takes into account the decrease of the LCHS success probability due to the decay of the simulated signal $\psi$ in time. 
This happens because of the dissipation in the considered nonunitary problem~\eqref{eq:initial-diff-equ}.

\ycb{For the optimal quadrature methods first proposed in this work and discussed further below, the quantum cost of constructing the weights, $Q_w$, is only weakly dependent on time.
Thus, it is subdominant to the cost of the selector.
In contrast, as will be explained further below, if trotterization is used~\cite{An23, Novikau24KvN} or if composite Gaussian quadrature is used~\cite{An23impr}, the cost of $Q_w$ increases superlinearly with time.
}

\subsubsection{Selector 
\ycr{Complexity} \ycb{with Efficient Block-Encoding}}\label{sec:selector-scaling}

The selector $S$ makes the main contribution to the cost of the LCHS algorithm, because $S$ scales with time.
In particular, since $S$ is implemented by using a QSP circuit, its query complexity in terms of the number of calls to the block-encoding oracle $U_C$ is 
\begin{equation}\label{eq:sel-cost-1}
    \oO(\|C_{\rm max} t\| + \log_2\varepsilon_{\rm QSP}^{-1}),
\end{equation}
where the simulated time $t$ is multiplied by the spectral norm $\|C_{\rm max}\|$ because  the encoded matrices $C_j$ are normalized according to Eq.~\eqref{eq:Cmax-norm}.
An important remark here is that $\|C_{\rm max}\|$ can grow linearly with $\kmax$.
To show this dependence explicitly, we rewrite Eq.~\eqref{eq:sel-cost-1} as
\begin{subequations}\label{eq:sel-cost-2}
\begin{eqnarray}
    &&\oO(\kmax\ybCm t + \log_2\varepsilon_{\rm QSP}^{-1}),\\
    &&\ybCm = \kmax^{-1}\|C_{\rm max}\|.
\end{eqnarray}
\end{subequations}
Here, according to Eq.~\eqref{eq:kmax-conv-sp}, the maximum value of the Fourier coordinate, $\kmax$, scales as 
\begin{equation}\label{eq:kmax-err-sp}
    \kmax = \oO(\yerr^{-1})
\end{equation}
if one uses the special kernel~\eqref{eq:xi-special}.
On the other hand, according to Eq.~\eqref{eq:err-scaling-impr-theory}, this scaling is improved exponentially,
\begin{equation}\label{eq:kmax-err-general}
    \kmax = \oO\left(\log^{\ycr{1/\beta}}\left(\yerr^{-1}\right) \right),
\end{equation}
if one applies the kernel~\eqref{eq:xi-general}.
\ycb{For approximate LCHS, the analytic approximations to the LCHS integral allow one to improve the truncation error to 
\begin{equation}\label{eq:kmax-err-optimal}
    \kmax = \oO\left( \log\left(\yerr^{-1}\right) \right).
\end{equation}
}

The QSP error increases with $N_k$, and, so, in order to have the total QSP error be close to the truncation error $\yerr$, the local error $\varepsilon_{\rm QSP}$ should scale at least as 
\begin{equation}
    \varepsilon_{\rm QSP} = \oO(\yerr/N_k).
\end{equation}
This results in the following complexity of the LCHS selector
\begin{equation}
    \oO\left(\kmax\ybCm t + \ycr{ Q_{\varepsilon,\rm QSP}} \right)
\end{equation}
\ycb{
where the work for QSP needed to achieve the target LCHS error is
\begin{equation} \label{eq:QSP_precision-goal}
Q_{\varepsilon, \rm QSP}:=\log_2\left(N_k \yerr^{-1} \right)=n_k+\log_2\left(  \yerr^{-1} \right).
\end{equation}
}

\ycb{Efficient block-encoding is essential for quantum advantage because, for problems with the right structure, this can exponentially reduce the classical complexity.}
If an efficient block-encoding of the matrices $B_{\rm m}$ and $A_H$ is provided, then each call to the oracle $U_C$ requires 
\begin{equation}\label{eq:Q-be}
    Q_{\rm BE} = \oO(n_k + \ycr{{\rm poly}(n_x, \varsigma^\prime)})
\end{equation}
gates where 
\ycb{$n_x = \log_2(N_x)$ is the number of qubits in the register $r_x$ required to store the solution vector on $N_x$ points of a spatial grid,} and the additive term $n_k$ appears due to the subcircuit~\ref{circ:sin} computing the sine function.

\ycb{
In our case of the linear advection-diffusion equation with constant coefficients (explored in Sec.~\ref{sec:results}), the matrices $A_H$ and $B_{\rm m}$ have a simple two and three-banded diagonal structure, respectively.
Hence, $\varsigma^\prime \leq 3$, with constant matrix elements, which can be easily encoded by several standard rotation gates, Fig.~\ref{circ:be-AH-AL}.
Therefore, the block-encoding error is defined by the precision of the standard rotation gates $R_y$ and $R_z$, which is the set to double floating-point precision in our numerical simulations.
For both matrices $A_H$ and $B_{\rm m}$, we take $\|M\| = \|C_{\rm max}\|$.
}

The scaling of the selector $S$ becomes
\begin{align}\label{eq:sel-scaling-final}
    Q_{\rm sel} = \oO\left(Q_{\rm BE}\left[\kmax\ybCm t + Q_{\varepsilon \ycr{,\rm QSP}}\right]\right)
\end{align}
For the final result, the dependence of $\kmax$ on $\yerr$ is described either 
by Eq.~\eqref{eq:kmax-err-sp}, 
Eq.~\eqref{eq:kmax-err-general}\ycb{, or by Eq.~\eqref{eq:kmax-err-optimal}}  depending on the chosen LCHS kernel.

\subsubsection{Comparison to Trotterization}
One can compare the \ycr{complexity} scaling of \ycb{the optimal} selector \ycb{in \eqref{eq:sel-scaling-final}} with the scaling of the selector based on the trotterization of $A_H$ and $A_L$ 
\ycr{employed} in our previous work~\onlinecite{Novikau24KvN}
\begin{equation}\label{eq:selector-complexity-prev}
    Q_{\rm trot} = \oO\left(Q_{\rm BE}\left[\|A_H t\| + \varepsilon_{\rm LCHS}^{-1} \|A_L t\| + Q_{\xi} \log_2\varepsilon_{\rm QSP}^{-1}\right]\right).
\end{equation}
In this case, the QSP approximation cost, $Q_{\xi}$, depends on the trotterization order, $p$, scaling as $t^{2 + 2/p}$,
\begin{equation}\label{eq:Q-xi}
    Q_{\xi} = \oO\left[
    \left(\frac{\|A_C\|\, \|A_L t\|^2}{\varepsilon_{\rm LCHS}^3}\right)^{1+1/p}\right].
\end{equation}
One can see that the complexity of our new selector, Eq.~\eqref{eq:sel-scaling-final}, does not include the multiplicative factor $Q_{\xi}$ which scales poorly with the simulation time. 
In addition, for time-independent problems, the algorithmic accuracy no longer depends on the trotterization step  
and, hence, on the commutator of the matrices $A_L$ and $A_H$.

\subsection{ Quadrature  Convergence Sets LCHS Complexity \label{sec:error-analysis}}
\ycb{In order to complete the analysis, we must select a numerical method to compute the LCHS integral and estimate $\Nk$ by placing bounds on the convergence of the error in computing the value of the integral.
First, we consider the prior art~\cite{An23,An23impr,Novikau24KvN} available at the time we first proposed our algorithm.
We analyze quadrature based on the trapezoidal rule~\cite{An23} in Sec.~\ref{sec:trapezoidal}, composite Gaussian quadrature~\cite{An23} in Sec.~\ref{sec:composite-Gauss}, and the trapezoidal rule after the $\sin(\theta)$ transformation~\cite{Novikau24KvN}.
Then, because this field is moving rapidly, in Sec.~\ref{sec:LS25}, we compare to recent results that consider approximate LCHS~\cite{Low25,Jin25}.
Perhaps, more importantly, in Sec.~\ref{sec:FCC} we consider how the improved convergence analysis for the trapezoidal rule based on the analyticity of the kernel~\cite{Low25} applies to the case of the near-optimal kernels \eqref{eq:xi-general}.
}

\ycb{
In Sec.~\ref{sec:FCC}, we prove that the sinusoidal transformation recasts the numerical integral as Fej\'er-Clenshaw-Curtis (FCC) quadrature\cite{Trefethen2008gauss} which can be computed classically using the fast Fourier transform (FFT).
Next, we consider the error convergence and the complexity in this setting.
We prove that, for kernels that are analytic on the real line, any numerical integration method that is based on interpolation rather than an exact projection generically has three asymptotic regions of convergence: an analytic region of exponential convergence as assumed in ~\onlinecite{Low25,Jin25}, a $p$-times differentiable region of power law convergence, and then a final region where convergence halts due to aliasing between nonlinear terms of different orders.
}

\ycb{
For the first time, our analysis definitively answers the question raised by L. Trefethen \onlinecite{Trefethen2008gauss}: \emph{Is Gauss[-Legendre] quadrature better than [Fej\'er-]Cleshaw-Curtis quadrature?}
{\bf The answer is no:}  both are Gaussian quadrature methods and both are matched optimally to different forms of the integrand.
In particular, FCC quadrature is optimally matched to integrands that are periodic and in $C^\infty$, i.e. smooth to all orders.
}

\ycb{
Then, we reconsider the analysis of \onlinecite{Low25} in light of the optimal convergence of FCC quadrature.
We prove that in the analytic region of exponential convergence, FCC quadrature is equivalent to Chebyshev–Gauss–Lobatto (CGL) quadrature, and, like any Gaussian quadrature method, converges twice as fast as when the integration points are not chosen in an optimal fashion.
Hence, for fixed $\kmax$, FCC converges better than the uniform sampling trapezoidal rule considered in \onlinecite{An23,An23impr,Low25} by a factor of 2.
While the rate of convergence is only improved by a constant of order unity, this is the best known constant and can essentially be considered optimal.
Moreover, the classical resources needed for the QSVT calculation based on FCC quadrature are optimal over other types of Gaussian quadrature because they can be computed with FFTs.
}

\ycb{Our analysis is the first to consider the multiple asymptotic regions that occur in practice as well as the impact of nonlinear aliasing between the two factors in the integrand: the weight function, $w(k)=\xi(k)/(1+ik)$, and the unitary, $e^{iC_jt}=e^{i(A_H+k_jA_L)t}$.
The estimates by other authors\cite{An23,An23impr,Low25} do not account for these facts, and, hence, may be too loose unless care is taken to resolve these issues.
First, it is important to understand how the choice of $\kmax$ impacts which of the three regions of asymptotic convergence that one should consider, as this will impact the dependence of complexity on $\Nk$.
Second,  it is important to understand that continuing to increase $\kmax$ can always improve convergence, and, that, when one is limited by classical resources, e.g. for computing and compiling the QSVT, one may very well need to choose $\kmax$ beyond the analytic region (1) of exponential convergence.
Third, we derive somewhat pessimistic lower bounds for complexity that are guaranteed to be correct.
Fourth, we point out that, since this part of the calculation must be performed classically, one can always estimate convergence by computing the FFT of the LCHS integral of interest.
}

\ycb{Finally, while our theoretical understanding of LCHS has greatly improved, our numerical results in Sec.~\ref{sec:results} have not been modified by the improved analysis.
The great virtue of our empirical numerical results is that they allowed us to carefully analyze and understand convergence -- proving that it was better than anticipated by the early analysis in \onlinecite{An23,An23impr,Novikau24KvN} -- before the improved analysis was developed.
}

\subsubsection{ Prior Art: Trapezoidal Rule \label{sec:trapezoidal}}
\ycb{First, consider the analysis of \onlinecite{An23}, which is based on using the} 
trapezoidal rule 
for the discretization of the LCHS integral~\eqref{eq:LCHS-theorem1}, 
the discretized step in the Fourier space should be small enough to make the discretization error close to the truncation error $\yerr$.
\ycb{The error in each subinterval is determined by the second derivative, which leads to the estimate 
\begin{align}
    \varepsilon^{\rm trap}_{\rm quad}= f''(k/\kmax)(2\kmax)^3/\Nk^2
\end{align} where $f(k/\kmax)$ is the integrand in terms of the scaled variable $k/\kmax$.}
This imposes the following condition on $N_k$ (see Sec. II in 
the Supplemental Material of Ref.~\onlinecite{An23}):
\begin{equation}\label{eq:Nk-cond}
   N_k = \oO\left(  \kmax^{3/2} \|A_Lt\|\ycr{m_\varepsilon^{1/2} \yerr^{-1/2}} \right).
\end{equation}
\ycb{
(Note that, in this equation, we have corrected a typo in the power law for $\yerr$ reported in \onlinecite{An23}, Sec. II, Eq.~S22.)
Clearly, reducing the dependence of $\kmax$ on $\yerr$ from Eq.~\eqref{eq:kmax-err-sp} to Eq.~\eqref{eq:kmax-err-general} or Eq.~\eqref{eq:kmax-err-optimal} will be beneficial in reducing the overall scaling. 
This yields a selector complexity that scales as
\begin{align} 
Q_{\rm sel} = \oO\left(Q_{\rm BE} \|\bar C_{\rm max}t\|\log^{1/\beta}(\yerr^{-1})\right).
\end{align}
} 

Now, let us consider the \ycb{complexity} scaling of the 
\ycr{calculation of the weights.}
As shown in \ycb{Eq. 90 in} Ref.~\onlinecite{Novikau24KvN}, if the oracle $U_{\sqrt{w}}$ computing the LCHS weights is implemented using QSVT circuits, then the \ycb{complexity} scaling of the weight computation including AA is
\begin{equation}\label{eq:w-scaling}
    Q_w = \ycb{\oO\left(n_k\left[\kmax^{3/2} \varepsilon_w^{-1}\right]\right)}
    =\oO\left(n_k\left[\kmax^{3/2} \ycb{m_\varepsilon} \yerr^{-1} \right]\right),
\end{equation}
assuming that the error of the weight computation is of the order of $\yerr$.
\ycb{While this has worse dependence on precision than the selector, this estimate will be remedied by the improved analysis in the next subsections. 
}

\subsubsection{ Prior Art: Composite Gaussian Quadrature = Trapezoidal Rule + Gauss-Legendre Quadrature
\label{sec:composite-Gauss}}
\ycb{
An, et al., ~\onlinecite{An23impr} analyzed the resources required by near-optimal LCHS based on composite Gauss-Legendre quadrature for numerical integration of the kernel.
This is equivalent to a standard $hp$-finite-element construction of the numerical integration rule, where the number of elements and the quadrature polynomial degree is set by the desired accuracy.
The truncation error can be bounded by 
\begin{align} \label{eq:error_trunc}
\varepsilon_{\rm trunc} &\leq \left[\int_{-\infty}^{-\kmax}+\int_{\kmax}^{\infty} \right]\abs{w(k)} dk
\\
&\leq m_\beta \int_{\kmax}^{\infty}e^{-k^\beta\cos(\beta\pi/2)}dk/k
\\
&=M_\beta E_1\left(k\cos^{1/\beta}(\beta\pi/2)\right)
\\
&\leq M_\beta e^{-\kmax^\beta \cos{(\beta \pi/2)}} \log{\left(1+\frac{1}{k\cos^{1/\beta}(\beta\pi/2)}\right)}.
\end{align}
where $M_\beta=m_\beta/\beta$.
The second line results from \onlinecite{An23impr} Appendix~D, Eq.~186, while the final line results from a bound for the exponential integral, $E_1(x)$.
Note that this is tighter than the bound reported in \onlinecite{An23impr} Lemma~9 Eq.~62.
Thus, in order to reduce the truncation error below threshold, one must choose
\begin{align} \label{eq:ktrunc_ACL}
\kmax \geq  \ktrunc:=\frac{\log^{1/\beta}\left(  M_\beta m_\varepsilon/\ktrunc \yerr\right)}{\cos^{1/\beta}{(\beta \pi/2)}} .
\end{align}
}

\ycb{
In order to reach the largest time scales without the Gibbs phenomenon ruining accuracy at the end points of the domain, the minimum value of $k$ must satisfy $\kmin\|A_L t\|<\pi$ where $\|\cdot\|$ is the spectral norm.\cite{Novikau24KvN}
For a numerical study of these issues, please see \onlinecite{Novikau24KvN} (Sec.~4.1 and Fig.~5(a)).
This yields the lower bound
\begin{align}
    \Nk:= 2\kmax/\kmin \geq \Ntrunc := 2\ktrunc \|A_Lt\|/\pi.
\end{align}
For the method of \onlinecite{An23impr}, they state it is sufficient to choose $\kmin \|A_L t\|\simeq 1$, and, hence, they use the more restrictive bound $\Nk\geq \pi \Ntrunc$.
Overall, $N_k$ scales as
\begin{align} 
N_k = \oO\left(  \|A_Lt\|  \log^{1/\beta}(\yerr^{-1})\right)
\end{align}
and the selector complexity scales as
\begin{align} 
Q_{\rm sel} = \oO\left(Q_{\rm BE} \|\bar C_{\rm max}t\|\log^{1/\beta}(\yerr^{-1})\right).
\end{align}
}

\ycb{
However, in order to control the quadrature error, \onlinecite{An23impr} (Sec.~3.1) estimated that a Gauss-Legendre quadrature rule of order 
\begin{align}\label{eq:nq_ACL}
    2\nq >  \log\left(\ktrunc  M_\beta'  m_\varepsilon \yerr^{-1} \right)
\end{align} 
is required, where $M'_\beta$ is another constant.
Here, the extra $\nq$ quadrature points are used to provide greater resolution within each of the $\Nk$ subintervals.
Thus, the total number of sample points, $\Nq = \nq \Ntrunc$, is  \onlinecite{An23impr} (Sec.~3.1) 
\begin{align} \label{eq:Nq_ACL}
   \Nq =  \oO{\left(  \ktrunc \frac{\|A_Lt\|}{\pi}  \log{\left(\ktrunc M_\beta'  m_\varepsilon\yerr^{-1}\right)} \right)}.
\end{align}
This is proportional to the cost of computing the answer classically.
}

\ycb{
As will be explained more fully in the next subsection, the overall work for the quantum composite Gaussian quadrature computation is only
\begin{equation}\label{eq:Qw_ACL_definition}
    Q_w:=\nk \Nq /p_w^{1/2}=\nk \nq \Ntrunc /p_w^{1/2},
\end{equation}
which yields
\begin{align} \label{eq:Qw_ACL}
    Q_w \sim 
   \oO{\left( \nk (2\ktrunc)^{3/2} \|A_Lt\|\log{\left(\ktrunc  M_\beta'  m_\varepsilon \yerr^{-1}\right)} \right)}
\end{align}
where, here, we use their bound $\Nk\geq\pi \Ntrunc$.
The extra $(2\ktrunc)^{1/2}$ increase in complexity over that reported in \onlinecite{An23impr} (Sec.~3.1) is due to the need to improve the success probability with AA.
The fact that  AA is necessary for this step was clearly demonstrated in our prior numerical studies \onlinecite{Novikau24KvN} and in the example below.
The overall scaling then becomes
\begin{align} 
Q_w = \bar \oO\left(\nk \|A_Lt\| \log^{1+3/2\beta}(\yerr^{-1})\right).
\end{align}
where the notation $\bar\oO$ indicates that this expression neglects
subdominant logarithmic factors.
}

\subsubsection{ Contemporary Results:  Trapezoidal Rule for Analytic Integrands \label{sec:LS25}}
\ycb{Low and Somma \onlinecite{Low25} present a detailed error analysis based on the two-step process of (1) applying uniform quadrature and (2) approximating the kernel with QSVT.
For the first step, their Lemma 10 proves that, when the kernel is analytic in a complex strip of height $|\Im(k)|<a$ and decays uniformly in the strip as $|k|\rightarrow\infty$, uniform quadrature leads to exponential convergence.
In this case, the error is bounded as $\yerr<M_a/(e^{2\pi a/k_{\rm \min}}-1)$, for some constant, $M_a<\hat M_a e^{a\|A_Lt t\|_1}$, where, in this context, $\|A_L\|_1=\|A_L\|_{L^1}$.
All kernels under consideration here, as well as in \onlinecite{An23,An23impr,Low25,Jin25} have poles at $k=- i$ and some also have poles at $k=i$; hence, one must choose $0<a<1$ for all of these cases.
Achieving a given error requires 
\begin{align}\label{eq:kmin-low25}
    \kmin&\leq2\pi a/\log\left(1+M_a \yerr^{-1}\right)<2\pi a/\log\left(M_a \yerr^{-1}\right)\nonumber\\
    &=2\pi /\left[\|A_Lt\|_1+ a^{-1} \log\left( \hat M_a \yerr^{-1}\right) \right],
\end{align}
and they make the choice $a=1/2$ in the final line.
Thus, the number of integration subintervals, $\Nk=2\kmax/\kmin$, 
\begin{align} \label{eq:Nk_LS}
\Nk 
= \kmax  \left[\|A_L t\|_1+ a^{-1} \log\left( \hat M_a \yerr^{-1}\right)\right]/\pi.
\end{align}
Due to the second term, this is somewhat larger than the estimate $\Nk\sim \oO(\kmax  \|A_L t\|)$ in the previous subsection.
Note, however, that, according to our study of the Gibbs phenomenon in \onlinecite{Novikau24KvN}, it would be more accurate to consider the periodic extension of the interval, which implies that one should replace $2\pi\rightarrow \pi$ in \eqref{eq:kmin-low25}, and this doubles the lower bound for $\Nk$ in \eqref{eq:Nk_LS}.
}

\ycb{
The other key difference is that, for the more optimal approximate LCHS kernel  described in \onlinecite{Low25} (Theorem 2), achieving the truncation error requires 
\begin{align} \label{eq:kmax_LS}
    \kmax\geq k_{\rm trunc}^{\rm approx-LCHS} =2c^{-1} \log{(M_c  \yerr^{-1} )}
\end{align}
for some constants $M_c$ and $c\neq 0$.
One can apply the integration strategy of \onlinecite{Low25} to the near-optimal kernel \eqref{eq:xi-general} by using the appropriate function.
This simply requires one to use the lower bound for  $\kmax$, given in \eqref{eq:ktrunc_ACL}, in the expressions above.
This yields the scaling
\begin{align} 
N_k = \oO\left( \left[ \|A_Lt\|+\log(\yerr^{-1})  \right] \log^{1/\beta}(\yerr^{-1})\right)
\end{align}
and a selector complexity that scales as
\begin{align} 
Q_{\rm sel} = \oO\left(Q_{\rm BE} \|\bar C_{\rm max}t\|\log^{1/\beta}(\yerr^{-1})\right).
\end{align}
}

\ycb{
For the second step, they provide a detailed estimate of the cost  of preparing the weights via a QSVT-style construction (see \onlinecite{Low25}  Sec.~5.2).
The cost scales as the product $Q_w=\nk \Nq/p_w^{1/2}$, where $\nk$ is the size of the $k$-register, $\Nq$ is the polynomial order required to achieve the desired quadrature error, and $p_w$ is the success probability of the block-encoding of the kernel function.
The analysis is dominated by the considerations of truncating the integral over the real line to a region of size $2\kmax$. 
The success probability can be shown to scale as 
\begin{align}
    p_w=\oO(1/2\kmax ),
\end{align}
and, hence, the complexity to amplitude amplify this to order unity is $p_w^{-1/2}=\oO(2\kmax)^{1/2}$.
We also confirmed this scaling in the numerical examples of \onlinecite{Novikau24KvN} and in the example below.
}

\ycb{
It can also be shown that, as long as the quantum state created has finite variance, the error in the state created by the QSVT weight evaluation is bounded by $\varepsilon_w<\oO((2\kmax)^{1/2}\hat\varepsilon_w )$, and, hence, the target QSVT error must satisfy 
\begin{align} \label{eq:error_QSVT}
\hat \varepsilon_w\leq \oO\left( \frac{\varepsilon_w }{(2\kmax)^{1/2}}\right) =\oO\left( \frac{\yerr}{m_\varepsilon(2\kmax)^{1/2}}\right).
\end{align} 
}

\ycb{
A significant improvement over the previous analysis is the recognition that, if the weight function is analytic, it can be approximated well with a globally convergent Chebyshev-Fourier series.
Hence, there is no need to compute multiple QSVT approximations for each subinterval.
Let the region of analyticity of the weight function,  $w(k)$, in the complex plane be specified by the Bernstein ellipse, $\Gamma$, 
\begin{align}
    k &=k_x+ik_y=\tfrac{1}{2}(\rho e^{i\theta}+\rho^{-1}e^{-i\theta}) \nonumber\\
   &= \tfrac{1}{2}(\rho+\rho^{-1}) \cos(\theta)+i\tfrac{1}{2}(\rho+\rho^{-1}) \sin(\theta).
\end{align}
In this case, the Bernstein ellipse is determined by 
\begin{align} \label{eq:rho-height}
    \rho-\rho^{-1}=2a/2\kmax = a/\kmax,
\end{align}
which has the solution
\begin{align}\label{eq:rho-solution}
    \rho=a/2\kmax+\sqrt{1+(a/2\kmax)^2}.
\end{align}
Then, the convergence of the Chebyshev series for $w(k)$ can be bounded via (\onlinecite{Trefethen2019approximation} Theorem 8.2)
\begin{align}
    \hat \varepsilon_w < M_\rho \rho^{1-\Nq}/(\rho -1),
\end{align}
where $M_\rho=4\max_{\partial \Gamma} (w)$ is determined by the maximum of $w(k)$ on the boundary of the ellipse, $\partial \Gamma$. 
This bound is reduced by $1/2$ if an exact projection is used instead of interpolation.\cite{Trefethen2019approximation}
Thus, this leads to the requirement
\begin{align} \label{eq:Nq_LS}
    \Nq-1 > \frac{\log{\left( M_\rho \hat\varepsilon_w^{-1}/[\rho -1] \right)}}{ \log{\left( \rho\right)}}.
\end{align}
Using the Taylor series approximation $\log(\rho^2)\simeq \rho^2-1+\dots$ near $\rho=1$ and the fact that $\rho-1\geq a/2\kmax$, it is sufficient to choose
\begin{align}
    \Nq-1 &\geq  \frac{2\kmax}{a} \log{\left( \frac{2\kmax M_\rho }{a \hat \varepsilon_w }  \right)}\\
          &= \frac{2\kmax}{a} \log{\left( \frac{(2\kmax)^{3/2} M_\rho' }{a \yerr }  \right)}
\end{align}
where, in the final line, we  define $M_\rho'= m_\varepsilon M_\rho$.
Thus, the overall scaling for $Q_w = \nk \Nq /p_w^{1/2} $ is 
\begin{align} \label{eq:Qw_LS}
    Q_w \geq  \oO\left(\nk \frac{(2\kmax)^{3/2}}{a} \log{\left(\frac{ (2\kmax)^{3/2}M_\rho'}{a\yerr}\right)}\right).
\end{align}
This strategy can be applied to the near-optimal kernel in Eq.~\eqref{eq:xi-general} by using the lower bound for $\kmax$ in Eq.~\eqref{eq:ktrunc_ACL}.
This yields a kernel complexity that scales as
\begin{align} 
Q_w = \bar\oO\left(n_k \log^{1+3/2\beta}(\yerr^{-1}) \right)
\end{align}
where the notation $\bar \oO$ indicates that this expression neglects subdominant logarithmic factors.
Here, we do not expand $n_k$ because it only depends logarithmically on $\|A_Lt\|$ and $\log(\yerr^{-1})$
Relative to Eq.~\eqref{eq:Qw_ACL}, the major improvement is that there is no factor of $\|A_L t\|$.
}

\ycb{
Note that, because the Chebyshev nodes do not agree with the uniform sample points used for the uniform integration method, the overall effective number of sample points is 
\begin{multline} \label{eq:Nq_Nk_LS}
    \Nq \Nk\sim \oO\left(\frac{2\kmax^2}{a}\left[\|A_Lt\|+ \frac{1}{a} \log(M_a\yerr^{-1}) \right] \right. \\
        \left. \times  \log{\left( \frac{2\kmax M_\rho' }{a \varepsilon_w }  \right)}\right).
\end{multline}
This represents the complexity scaling of the classical algorithm for computing the integral with this method.
Note that this scales similarly to $\Nq$ in Eq.~\eqref{eq:Nq_ACL} for the method of the previous subsection,\cite{An23impr} but has worse leading dependence on $\kmax$, because information is not reused between the uniform sample points for the integral and the Chebyshev nodes for the weights.
}

\subsubsection{ Our Method: Fej\'er-Clenshaw-Curtis Quadrature \label{sec:FCC}}
\ycb{Now, let us consider the error convergence for our approach.
Using the sinusoidal transformation in Eq.~\eqref{eq:coord-transf} to perform the truncated integral over the real line is equivalent to the use of Fej\'er-Clenshaw-Curtis (FCC) quadrature over the interval $\theta\in[-\pi/2,\pi/2)$. 
Thus, we do not need to introduce an additional Gaussian quadrature at each step as in \onlinecite{An23impr}.
FCC quadrature is known to have exceptionally good convergence properties, often similar to or even rivaling that of Gaussian quadrature.\cite{Trefethen2008gauss}
While much has been written about the mysterious virtues of FCC,\cite{Trefethen2008gauss} a clear explanation has not been provided until now.
Here, we prove that there is a simple explanation for why FCC quadrature performs so well in practice: for a periodic and analytic integrand, it is equivalent to CGL quadrature with the integrand multiplied by $\cos(\theta)$.
Thus, relative to \onlinecite{An23impr}, our approach clearly reduces the overall complexity of the algorithm as well as the resources required. 
}

\ycb{
The uniform discretization of the angle in Eq.~\eqref{eq:coord-transf-disc} can be interpreted as the evaluation of a Chebyshev quadrature formula over the integrand $dk$ rather than $dk/(1-(k/\kmax)^2)^{1/2}$, which is equivalent to Chebyshev quadrature of the integrand multiplied by $(1-(k/\kmax)^2)^{1/2}$.  
In turn, this is equivalent to the sum of a Fourier series of the integrand multiplied by $dk/d\theta=\kmax\cos(\theta)$.
Hence, FCC quadrature simply uses ``uniform quadrature'' for the Fourier series, which corresponds to using Chebyshev nodes for the quadrature formula.
Given the Fourier series, $f(\theta)=\sum_{m=-\infty}^\infty \hat f_m e^{im\theta}$, the FCC quadrature rule can be computed analytically
\begin{align} \label{eq:FCC-quad}
 \int_{-\pi/2}^{\pi/2} f(\theta)\cos{(\theta)} d\theta
 &=\sum_{m=-\infty}^\infty \frac{\hat f_{2m}}{2i(m+1)}+\frac{\hat f_{2m}}{2i(m-1)}
 \\
 &=\sum_{m=-\infty}^\infty (-1)^{m-1}\frac{ 2\hat f_{2m}}{(2m)^2-1}.
\end{align}
A significant advantage of this method is that it can be performed rapidly via fast Fourier transform (FFT) methods\cite{Gentleman1972i,Gentleman1972ii};
one could even consider the quantum Fourier transform (QFT).~\cite{Chen2023quantum}
}

\begin{lemma}[\bf FCC quadrature $\equiv$ Fourier quadrature $\equiv$ CGL quadrature] \label{thm:FCC=CGL}
For the class of functions, $f(\theta):\Reals\rightarrow \Reals$, that are periodic on $\theta\in[0,2\pi)$ with convergent Fourier series, FCC quadrature of the integrand $\int_{-\pi/2}^{\pi/2} f(\theta)\cos(\theta) d\theta$ is equivalent to integration of the integrand, $f(\theta)\cos(\theta)$, by Fourier series.
Applying the coordinate transformation, $z=\sin{(\theta)}$, proves that both are equivalent to CGL quadrature of the integrand $f(\arcsin(z))(1-z^2)^{1/2}$, i.e. the integral $\int_{-\pi/2}^{\pi/2} f(\arcsin(z))dz$, with nodes at the Chebyshev extrema, which are the standard discrete Fourier series nodes.
Hence, FCC convergence is set by Fourier series convergence and is equivalent to CGL convergence. 
Thus, for an $N$-point method, Fourier harmonics in $\theta$ and polynomials in $z$ are integrated exactly up to order $2N-1$.
\end{lemma}
\begin{proof}
Using the change of variables $z=\sin(\theta)$ proves that the continuous integrals
\begin{align}
    \int_{-\pi/2}^{+\pi/2} f(\theta) \cos(\theta) d\theta = \int_{-1}^{+1} f(\arcsin(z))  dz
\end{align}
are equal.
Using this same change of variables proves that the two discrete approximations to these integrals, FCC quadrature (left) and   CGL quadrature (right): 
\begin{align} \label{eq:FCC-quad-rule}
   \frac{1}{N} \sum_j f(\theta_j) \cos(\theta_j)  = \frac{1}{N}\sum_j f(\arcsin(z_j)) (1-z_j)^2,
\end{align}
are equal.
This coordinate transformation also proves that FCC quadrature and CGL quadrature at the extrema use the same nodes and weights.
Because the Chebyshev-Fourier series for $f(\theta)$ converges, the Chebyshev-Fourier series for $F(\theta) :=\int_{-\pi/2}^\theta f(\theta') d\theta'$ converges
and, hence, the integral $F(\pi/2)$ exists and is finite.
Convergence estimates follow from the theory of Gaussian quadrature.\cite{Trefethen2019approximation}
\end{proof}

\ycb{
The importance of FCC quadrature is that it is optimally adapted to the Chebyshev series used by QSVT.
Thus, it can directly use data at the Chebyshev extrema and can converge more rapidly than the uniform quadrature method in \onlinecite{Low25}, which is not matched in an optimal manner.
While the form of the LCHS integral used in \onlinecite{An23impr,Low25} is better matched with Gauss-Legendre quadrature, it is not usually recommended to go to exceptionally high order because the numerical calculations of the nodes and weights require exceptionally high precision.~\cite{Trefethen2019approximation}
The need to avoid this issue is implicitly recognized in \onlinecite{An23impr} with their choice of composite Gaussian quadrature.
In contrast, the FCC quadrature used here does not require an approximate numerical computation of the nodes and weights because it is equivalent to integration by Fourier series, and, hence, can be used to arbitrarily high order.
}

\ycb{
The accuracy of the kernel function is determined by the accuracy of the QSVT approximation.
If the Chebyshev expansion of the kernel is known in closed form, then one has an accurate expression for the Chebyshev coefficients; i.e. using the projection onto Chebyshev polynomials.
However, for generic kernels such as \eqref{eq:xi-general}, the Chebyshev expansion is not known in closed form and the Chebyshev coefficients must be evaluated numerically through interpolation rather than projection.
This implies that there is an additional aliasing error due to the fact that the interpolants cause aliasing of polynomial orders that are higher than the finite range of orders included in the calculation.
}

\ycb{
In this case, the kernels of interest, $w(k)=\xi(k)/(1+ik)\in L^1$, are meromorphic, analytic on the real line, and decay as $k\rightarrow\pm\infty$.
However, once the integral is truncated to the finite domain $k\in[-\kmax,+\kmax)$, the periodic extension of $w(k)$ is no longer smooth to all orders at the boundary.
For example, both the Cauchy kernel and the real part of the near-optimal kernels are continuous but not smooth at the boundary, i.e. they are in $C^1$, and  
the imaginary part of the near optimal kernel is discontinuous at the boundary, i.e. it is in $C^0$.
Even though the optimal approximate LCHS kernels\cite{Low25, Jin25} are analytic, i.e. in $C^\omega$, after truncating the integral to finite $\kmax$, the periodic extension of these kernels is no longer in $C^\omega$ and the same issues arise.
Yet, thanks to the decay condition, as $\kmax\rightarrow \infty$, the ``part'' of $w(k)$ that is not smooth becomes very small relative to the ``part'' that is analytic.
Since the Fourier series for the kernel is the sum of two parts that have different asymptotics, the sum will generically display multiple asymptotics.
These considerations are formalized in the following lemma:
}

\begin{lemma}[\bf Discrete Fourier series can have multiple asymptotics] \label{thm:multiple-asymptotics}
Consider a function, $w(\theta):\Reals\rightarrow\Reals$, that is periodic on $\theta\in [0,2\pi)$, $p$-times differentiable, $w\in C^p$, and has a convergent Fourier series  $w(\theta)=\sum_{m=-\infty}^\infty \hat w_m e^{im\theta}$.
Assume the function of interest has the form $w=w_{\rm \infty}+w_p $, where: (1) $w_{\rm \infty},w_p$ are periodic and have convergent Fourier series; (2)  $w_{\rm \infty}\in C^\infty$ is smooth to all orders;   and (3) $w_{\rm p}\in C^p$ is $p$-times differentiable.
The bandwidth-limited discrete Fourier series coefficients  
\begin{align}
\hat w_m:=\Nq^{-1}\sum_{j=0}^{Nq-1} w(\theta_j)e^{ i m \theta_j},
\end{align}
where $\theta_j=2\pi j/N$, can have at least three asymptotics:
\begin{enumerate}
\item First, a region of faster than polynomial decay generated by the part of  $w(\theta)$ that is in $C^\infty$.
\item Second, a region of power law decay of the form $m^{-(p+1)}$ generated by the part of $w(\theta)$ that is in $C^p$ and not in $C^\omega$.
\item Third, a region where aliasing completely halts convergence at the Nyquist frequency. 
\end{enumerate}
\end{lemma}
\begin{proof}
The sum of the Fourier series coefficients $\hat w_m=\hat w_{\infty,m} + \hat w_{p,m}$ has three components with different asymptotics:
\begin{enumerate}
\item Faster than polynomial decay generated by the infinitely-smooth component $w_\infty\in C^\infty$.
\item Power law decay of the form $m^{-(p+1)}$ generated by the $p$-times differentiable component $w_p\in C^p$.
\item Aliasing error, due to all harmonics that are higher than the maximum $N$ retained by DFT, halts convergence at the Nyquist frequency. 
\end{enumerate}
Depending on the relative size of $w_\infty$ and $w_p$, as well as on the resolution, $\Nq$, one or more of these asymptotics will be present and observable.
For the assumption $\|w_\infty\|_1 \gg \|w_p\|_1$, all three asymptotics are present. 
If $w_\infty$ is actually analytic, $w_\infty\in C^\omega$, then region (1) exhibits exponential decay.
\end{proof}

\begin{figure}[t]
\centering
\includegraphics[width=3in]{ 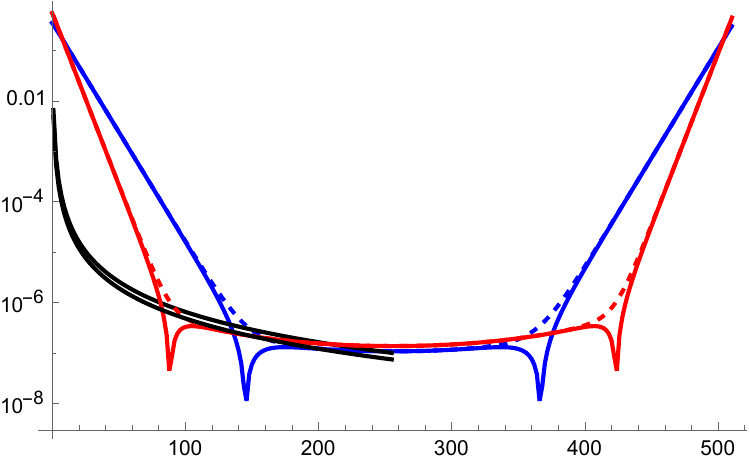}\\
\includegraphics[width=3in]{ 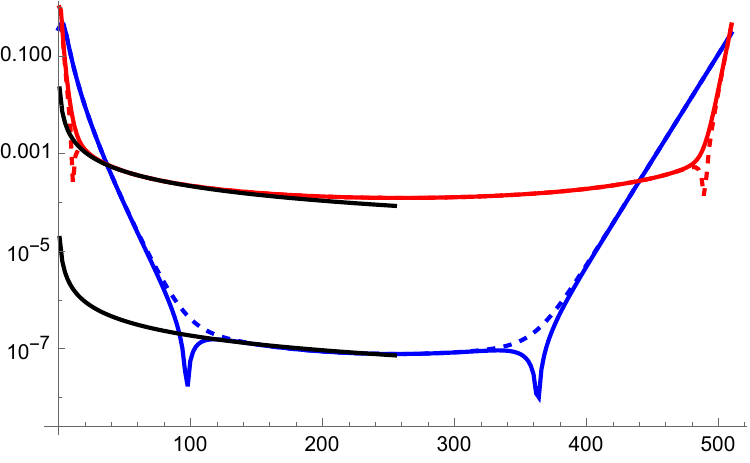}
\caption{
Absolute value of Chebyshev-Fourier coefficients vs. $j$ before (red) and after (blue) the $\sin(\theta)$ transformation for the:
(a) Cauchy kernel~\eqref{eq:xi-special} and
(b) near-optimal kernel~\eqref{eq:xi-general}.
Both plots use parameters $\beta=0.75$, $\kmax=20$, $\nk=9$, $\Nk=512$ and show the the even $j$ coefficients (solid) and the odd $j$ coefficients (dashed) separately.
For all cases, the initial exponential decay eventually turns into a power law decay, then halts completely due to aliasing.
Fits to the form $m^{-p}$, where (a) $p=2$ or (b) $p=1$ are shown as a guide to the eye (black, solid). 
}
\label{fig:kernel_fft}
\end{figure}

\ycb{
These three different asymptotics are clearly illustrated in Fig.~\ref{fig:kernel_fft}.
Clearly, it is best to use the Chebyshev expansion in the region where it is converging exponentially quickly.
If higher accuracy is desired, then it is best to add more quadrature nodes; i.e. increase $\Nq$.
}

\ycb{
For the near-optimal kernels \eqref{eq:xi-general} on the periodic domain $k\in[-\kmax,\kmax)$, the power law in region (2) is determined by the fact that the real part has a discontinuous derivative, so that $p=1$ and the coefficients scale as $m^{-2}$, and the fact that the imaginary part is itself discontinuous, so that $p=0$ and the the coefficients scale as $m^{-1}$.
Because applying the FCC quadrature rule improves convergence by the factor $m^{-2}$, for the coefficients the sum \eqref{eq:FCC-quad-rule}, this improves to $m^{-4}$ for the real part and $m^{-3}$ for the imaginary part.
For the Cauchy kernel \eqref{eq:xi-special}, this region has the same scaling as the real part of the near-optimal kernel.
The more optimal kernels proposed in \onlinecite{Low25, Jin25} will have scaling similar to these examples.
}

\ycb{
It is important to point out that the estimates provided by other authors have not included the asymptotic regions of reduced convergence: (2) and (3).
Therefore, the complexity analysis~\cite{An23impr,Low25,Jin25} may be compromised if one is not careful to ensure that $\kmax$ stays within region (1).
Clearly, one must take care in ensuring that the power law ``noise floor'' set by regions (2) and (3) are sufficiently well suppressed to use the estimates that correspond to the analytic part $w_\infty$.
Yet, for a practical calculation, it would be a mistake to not include regions (2) and (3) when necessary because these regions can always reduce the error.
The only thing these regions affect negatively is the complexity analysis and this may not matter for a practical application where one is limited by resolution requirements.
This one of the reasons why it is important to perform numerical studies of convergence.
}

\ycb{
The $\cos(\theta)$ factor has the benefit of making the integrand vanish more quickly at the endpoints $\theta=\pm\pi/2$.
In turn, this pushes the error floor, due to non-smooth behavior of the integrand, for a given $\Nk$ much further down.
Fig.~\ref{fig:kernel_fft}(b) shows that this effect is more pronounced for the near-optimal kernel.
Unfortunately, the fact that the cosine has the form of a square root, i.e. $\cos(\theta)=(1-\sin^2(\theta))^{1/2}$, tends to reduce the rate of convergence by a factor of $\oO(1)$, which then requires higher $\kmax$.
In our numerical studies, we found that, to achieve the same error in approximating the weight function, $\kmax$ needs to increase by a factor of $\sim 2$.
However, because our method uses FCC quadrature for quadrature of the LCHS integral, this doubles the rate of convergence by a factor of 2.
Thus, there is no actual cost increase, and, for the near-optimal kernels, it appears there is a net benefit.
Even if the overall effect was to increase the cost of computing the weights, this has negligible impact on the cost of the selector, which is the most important cost of LCHS.
This is because, the part of the query complexity of the selector that is linear in time scales as $\nk=\log_2(\Nk)=\log_2(2\kmax/\kmin)$, and, hence only increases logarithmically with $\kmax$.
}

\ycb{
In the region where FCC quadrature is dominated by the smooth component $w_\infty$, the convergence rate is controlled by CGL quadrature, i.e. improved by a factor of 2 relative to non-Gauss-type methods such as the trapezoidal rule.
For $\Nq$-point CGL quadrature of a function, $w$, analytic within the Bernstein ellipse $\Gamma$ of radius $\rho$, which is exact for polynomials of order $2\Nq-1$, the convergence rate is \cite{Trefethen2019approximation}
\begin{align}
\hat \varepsilon_w \leq M_{\rho^2} \rho^{2(1-\Nq)}/(\rho^2 -1),
\end{align}
where $M_{\rho^2}=5\max_{\partial\Gamma} (w)$ is determined by the maximum on the boundary of the ellipse.
Thus, this leads to the requirement
\begin{align}
\Nq-1 \geq 
  \frac{\log{\left( M_{\rho^2} \hat \varepsilon_w^{-1}/[\rho^2 -1] \right)}}{ \log{\left( \rho^2\right)}}.
\end{align}
}

\ycb{
Considering the fact that there are multiple regions of asymptotic convergence, one must take care in the analysis when $\kmax>\ktrunc$.
If we assume analytic convergence within the region of size $\ktrunc$, this same rate of convergence may not hold up to $\kmax$.
Yet, because the error can always be improved by increasing $\kmax$, we only demand analyticity within a region of size $\abs{\Re{(k)}}\leq \ktrunc$.
This is equivalent to replacing $a/\kmax\rightarrow a/\ktrunc $ in Eqs.~\eqref{eq:rho-height} and \eqref{eq:rho-solution}.
Hence, the rescaled Bernstein ellipse is determined by 
\begin{align}\label{eq:rho2-height}
    \rho -\rho^{-1}=2a/2\ktrunc = a/\ktrunc,
\end{align}
which has the solution
\begin{align}\label{eq:rho2-solution}
    \rho=a/2\ktrunc+\sqrt{1+(a/2\ktrunc)^2}.
\end{align}
Using the Taylor series approximation $\log(\rho^2)\simeq \rho^2-1+\dots$ near $\rho=1$ and the fact that the Bernstein ellipse satisfies the relation $\rho^2-1\geq  a/\ktrunc$, implies that it is sufficient to choose
\begin{align} \label{eq:Nq_FCC}
    \Nq &\geq 1+\frac{\ktrunc}{a}\log{\left(\frac{ \ktrunc M_{\rho^2} }{a \hat \varepsilon_w }\right)}
       \\ 
       &= 1+\frac{\ktrunc}{a}\log{\left(\frac{  (2\kmax)^{1/2} \ktrunc M_{\rho^2}' }{a\yerr }\right)}
\end{align}
where, in the final line, $M_{\rho^2}'=m_\varepsilon M_{\rho^2}$.
To derive this formula, we use the same error requirement as before, $\varepsilon_{\rm QSVT} \leq \yerr/m_\varepsilon(2\kmax)^{1/2}$, because  this error  is determined by the full range of $\pm\kmax$.
Clearly, due to the quadratically improved rate of convergence of CGL quadrature over uniform sampling, this method has better scaling than Eq.~\eqref{eq:Nq_LS} by a factor $\gtrsim 2$.
}

\ycb{
Once again, in order to reach the largest time scales via FCC/FFT, we must impose the requirement~\cite{Novikau24KvN} 
\begin{align} \label{eq:kmin_FCC}
    \kmin\|A_Lt\|= \pi
\end{align}
and, hence, the requirement 
\begin{align}
    \Nk :=\frac{2\kmax}{\kmin} \geq \Ntrunc  :=\frac{2\ktrunc}{\kmin}= 2\ktrunc \frac{\|A_L t\|}{\pi}.
\end{align}
In order to prevent aliasing between the computation of the weights, $w(k)$, and the Hamiltonian simulation of $e^{ikA_Lt}$, for the FCC/FFT method, we must sum the requirements for both, so that $\Nk=\Nq+\Ntrunc$. 
Thus, we arrive at the conclusion
\begin{align} \label{eq:Nk_FCC}
   \Nk > \ktrunc\left[\frac{2\|A_L t\|}{\pi} 
        + \frac{1}{a} \log{\left(\frac{(2\kmax)^{1/2}\ktrunc  M_{\rho^2}' }{a\yerr } \right)}\right].
\end{align}
We note that previous reported lower bounds for $\Nk$ may be too loose due to aliasing issues.
}

\ycb{
The bound for $\Nk$ can be rephrased in terms of $\kmax$ as
\begin{align} \label{eq:kmax_FCC}
   \kmax  > \ktrunc \left[1
        + \frac{\pi}{2a \|A_Lt\|} \log{\left(\frac{(2\kmax)^{1/2}\ktrunc  M_{\rho^2}' }{a\yerr }\right)}\right].
\end{align}
Because our results strictly require $\kmax>\ktrunc$, for FCC quadrature, the extra resolution is always used to increase the range of $k$ and, hence, has the benefit of improving the overall weight, quadrature, and truncation accuracy simultaneously.
This is rather different than the quadrature methods of the previous subsections,~\cite{An23,An23impr,Low25} which required subsampling to yield higher resolution within each of the $\Nk$ subintervals.
For those methods, subsampling was needed to generate high enough accuracy for either the weights or the quadrature error, but did not improve other factors such as the truncation error.
}

\ycb{
In the final form of \eqref{eq:kmax_FCC}, we see that it was important to use $\abs{\Re{(k)}}\leq \ktrunc$ as the region of analyticity, rather than $\abs{\Re{(k)}}\leq \kmax$, otherwise this requirement could present a contradiction for sufficiently low $\yerr$.
Had we used $\kmax$ instead  of $\ktrunc$, we would have arrived at the relation
\begin{align}  
   \kmax  >  \ktrunc
        + \frac{\pi \kmax}{2a \|A_Lt\|} \log{\left(\frac{2^{1/2}\kmax^{3/2}   M_{\rho^2}' }{a\yerr }\right)}.
\end{align}
This leads to the requirement
\begin{align}  
   \kmax  >   \ktrunc 
           \left[1- \frac{\pi }{2a \|A_Lt\|} \log{\left(\frac{2^{1/2}\kmax^{3/2}   M_{\rho^2}' }{a\yerr }\right)}\right] ^{-1}.
\end{align}
which is only reasonable if the second factor is positive.
In turn, this provides a restriction on the requested precision
\begin{align}  
\frac{ 2\|A_L t\|}{\pi}  >  
  \frac{ 1}{a} \log{\left(\frac{2^{1/2}\kmax^{3/2}   M_{\rho^2}' }{a\yerr }\right)} .
\end{align}
Because the right hand side only increases logarithmically, this inequality may hold true at times that sufficiently large.
However, this inequality will always fail for short times at fixed precision.
Nevertheless, the result \eqref{eq:kmax_FCC} is always valid.
}

\ycb{
With the FCC method, the total work for preparing the weights quantumly, $Q_w   =\nk\Nq/p_w^{1/2}$, takes the simple form 
\begin{align} \label{eq:Qw_FCC}
    Q_w &  \sim \oO\left(
        \nk (2\kmax)^{1/2} \frac{\ktrunc}{a}\log{\left(\frac{  (2\kmax)^{1/2} \ktrunc M_{\rho^2}' }{a\yerr }\right)}
    \right).
\end{align}
This is because the success probability still depends on the total range of $\pm\kmax$, so that $p_w=1/2\kmax$.
For a classical pseudo-spectral method, one would compute the weights and the nodes from the Chebyshev-Fourier series, multiply the two factors of the integrand in real space, perform the FCC quadrature with the final Chebyshev-Fourier series, and then evaluate the result in real space.
Classically, it is efficient to use FFTs and to place the weights and nodes on the same grid of size $\Nk$.
Thus, the classical complexity is $3\Nk \log(\Nk)$ for the three FFTs and $3\Nk$ arithmetic operations: $\Nk$ multiplications to compose the integrand from the two factors, $\Nk$ divisions to perform the quadrature rule, and a final sum of $\Nk$ terms to compute the integral.
}

\ycb{
Quantumly,  
the FCC complexity in \eqref{eq:Qw_FCC} is better than the trapezoidal rule of \eqref{eq:Qw_LS}.
This is because the factors of $(2\kmax)^{3/2}$ in \eqref{eq:Qw_LS} are replaced with factors of $(2\kmax)^{1/2}\ktrunc$ in \eqref{eq:Qw_FCC}, so the main overall effect is an improvement in complexity by the factor $2\kmax/\ktrunc$.
Even for fixed $\kmax\sim \ktrunc$, the complexity is $2\times$ better than \eqref{eq:Qw_LS}.
All of these results are improved over \eqref{eq:Qw_ACL} because the preparation steps do not scale with time $\oO(\|A_L t\|)$.
}

\ycb{
It is also notable that the classical complexity scaling of an FCC quadrature-based pseudospectral method, $\sim \oO(3\Nk\log{(\Nk))}$, is better than the results of the previous quadrature methods.
Thus, for the practical example studied here, at first, the coordinate transformation appears to require an $\sim 2\times$ larger $\kmax$, but this cost increase is then paid back by the use of FCC quadrature, which results in $2\times$ faster convergence of the error in the numerical quadrature rule.
Even if this were not the case, the QSVT cost, $Q_w$, is typically subdominant to the cost of the selector, $Q_{\rm sel}$, which increases with time.
}

\ycb{
In terms of precision, the overall selector complexity can be summarized as 
\begin{multline}\label{eq:Qsel_FCC_precision}
Q_{\rm sel}=\oO\left(Q_{\rm BE} \|\bar C_{\rm max}t\|\log^{1/\beta}(\yerr^{-1})  \right.
\\
\left. \times \left[1+\frac{\pi}{2\|A_Lt\|} \log(\yerr^{-1})\right]\right).
\end{multline}
For example, for the new optimal kernels where $\beta=1$, the polynomial in this expression is quadratic in $\log(\yerr^{-1})$ but the impact of this term decays in time.
The overall kernel complexity can be stated as
\begin{multline} \label{eq:Qw_FCC_precision}
Q_w=\bar \oO\left( n_k   \log^{1+3/2\beta}(\yerr^{-1})\right.
\\
 \left. \times  \left[1+\frac{\pi}{2\|A_Lt\|} \log(\yerr^{-1})\right]\right),
\end{multline}
neglecting subleading logarithmic factors.
Again, we remind the reader that $n_k$ only depends logarithmically on $\|A_Lt\|$ and $\log(\yerr^{-1})$.
Finally, we arrive at a new theorem for the FCC-based LCHS algorithm.
}
\begin{theorem}[{\bf FCC-LCHS Algorithm}] \label{thm:FCC-LCHS}
{ There is a quantum LCHS algorithm based on integration of either an optimal approximate~\cite{Low25,Jin25} or near-optimal exact~\cite{An23impr} kernel using FCC quadrature that prepares the normalized solution of Eq.~\eqref{eq:initial-diff-equ} with $\Omega(1)$ success probability and a flag indicating success that uses $Q_A$ queries to the oracle encoding $A$ and $Q_{in}$ queries to the initial state preparation oracle, where}
\begin{align}
    Q_A&= \oO\left( \frac{\norm{\psi(0)}}{\norm{\psi(t)}} \|A t\| \log^{1/\beta}(\yerr^{-1})\right)
    \\
    Q_{in} &= \oO\left( \frac{\norm{\psi(0)}}{\norm{\psi(t)}}\right).
\end{align}
The parameters are set via the equations: \eqref{eq:kmin_FCC} for $\kmin$,  \eqref{eq:Nk_FCC} for $\Nk$, \eqref{eq:kmax_FCC} for $\kmax$.
The gate complexity of the selector is $Q_{\rm sel}$ in \eqref{eq:Qsel_FCC_precision} and the gate complexity for the kernel weights is $Q_w$ in \eqref{eq:Qw_FCC_precision}.
If an optimal kernel is used, as in \onlinecite{Low25,Jin25}, then $\beta=1$ above and $\ktrunc$ is set by  \eqref{eq:kmax_LS}.
If a near-optimal kernel is used, as in \eqref{eq:xi-general}, then $\ktrunc$ is set by \eqref{eq:ktrunc_ACL}.
\end{theorem}
\begin{proof}
The LCHS integral can be approximated optimally using LCU~\cite{Childs12} where the weight functions are computed optimally using QSVT\cite{Low19} and the success probability is boosted with amplitude amplification~\cite{Brassard02}. 
Using the sinusoidal coordinate transformation in \ref{eq:coord-transf-disc}, the LCHS integral is computed using the FCC quadrature rule.
Given this transformation and an efficient block-encoding of $A_H$ and $A_L$, the selector can be performed optimally by using a single QSP and qubitization step to compute all required Hamiltonian simulations.
The explicit circuit is described in Sec.~\ref{sec:circuit} and the complexity analysis is completed in Sec.~\ref{sec:FCC}.
\end{proof}
 
\subsection{Further Considerations}\label{sec:further-considerations}
\subsubsection{FCC reduces memory \& cost relative to number operator  \label{sec:ancilla}}
\ycb{
Another point to consider is that Low \& Somma \onlinecite{Low25} use a diagonal ``number operator,'' $D=\sum_k k\ket{k}\bra{k}$, construction for the $O_{\sqrt{w}}$ (PREP) operators that is similar to the one they use for the selector $S$.
Relative to the $\sin(\theta)$ encoding, this procedure requires an additional $\nk$ ancillary register and a SWAP of the two registers which costs $\oO( 3 \nk)$ extra 2-qubit gates.
If all of these extra ancillary registers can be reused, then the total increase in memory is $2\nk$ and costs at least an  additional $2\times \oO(3\nk)$ additional 2-qubit gates.
However, if none of these registers can be reused, this requires an ancillary register of size $4\nk$ and costs at least an additional $4\times \oO(3 \nk)$ 2-qubit gates.
We believe that that the $O_{\sqrt{w}}$ (PREP) operators might be able to reuse the extra ancillary register but that the selector cannot, which only requires an intermediate size register $3\nk$ and $3\times \oO(3\nk)$ additional 2-qubit gates.
Furthermore, at the end of Sec.~\ref{sec:results}, we note that this also increases the number of elementary gates needed to represent the single-target-multi-control (STMC) gates that appear in the circuit.
}

\ycb{
To compare to our approach, even if the coordinate transformation slowed convergence by $4\times$, so that after accounting for the $2\times$ faster convergence of the quadrature error,  one would still require $\Nk$ to be larger by $2\times$, this means that  $\nk$ is only larger by a single qubit.
}

\ycb{
While these considerations do not change the asymptotic complexity, in practice, it is important to reduce the size of the ancillary register, and, this is one of the benefits of the sinusoidal transformation approach.
For the simple example studied here, we found $\nk\approx 8 $ offered good resolution, but the approach of \onlinecite{Low25} might require an ancillary register of 16-32 qubits, which means the total required memory would no longer fit on a single GPU. 
Near-term quantum computers must also avoid large ancillary registers whenever possible.
}

\subsubsection{Remarks on Schr\"odingerization \label{sec:Schrodingerization}}
\ycb{For completeness, we point out that, according to the appendices of \onlinecite{Low25},  Schr\"odingerization effectively uses an asymmetric version of LCU of the form
$Sch=O_{\rm left}^\dagger S O_{\rm right}$, such that $O_{\rm left, right}$ prepares a block-encoding of $f_{\rm left,right}$  subject to the condition $ f^*_{\rm left} (k)f_{\rm right}(k) = w(k)$. 
This is suboptimal to the symmetric encoding used in the usual LCU form, $O_{\sqrt{w}}$, because the complexity grows with the subnormalization of the block-encoding.
One can prove this using the fact that, for an asymmetric encoding, the cost of $Q_w$ is controlled by the product of the two norms $\|f_{\rm left}\|_2 \|f_{\rm right}\|_2$.
The cost is minimized for the symmetric encoding, where the norms satisfy $\|f_{\rm left}\|_2  \|f_{\rm right}\|_2=\|w\|_1 $.
}

\ycb{
Yet, we also note that, when using either our method or the method of Ref.~\onlinecite{Low25}, the cost increase for $Q_w$ from the QSVT step that constructs the weights is typically benign relative to the cost of the selector, because the cost of the latter must grow with time.
If one uses the integration method of Ref.~\onlinecite{An23impr} for Schr\"odingerization, the complexity analysis in Sec.~\ref{sec:composite-Gauss}, shows that $Q_w$ also grows with time, so it becomes important to consider the optimal form of LCHS.
As mentioned in Sec.~\ref{sec:complexity}, the same issue occurs if one must use trotterization and this is discussed more fully in ~\onlinecite{An23impr,Novikau24KvN}.
}

\section{Numerical simulation results}\label{sec:results}

To investigate the scaling and the success probability of the described LCHS circuit, we simulate the advection-diffusion equation (ADE):
\begin{equation}\label{eq:ADE}
    \partial_t\psi = - v\partial_x\psi + D\partial_x^2 \psi
\end{equation}
with a uniform velocity $v = 1.0$ and diffusivity $D = 0.01$.
The detailed quantum circuit for modeling this equation is given in Ref.~\onlinecite{code-OPT-LCHS}.
The initial condition $\psi(0,x)$ is a Gaussian centered at $x = 0.5$ with the width $0.05$.
The simulated domain, $x = [0,1)$, with periodic boundary conditions is discretized with $N_x$ spatial points.
After the discretization, Eq.~\eqref{eq:ADE} is recast as Eq.~\eqref{eq:initial-diff-equ} with the following matrix:
\begin{equation}\label{eq:ADE-A}
    A_{i_r i_c} = -
    \left\{ \begin{aligned}
        c_{-1}&,\quad i_c = i_r - 1,\\
        c_{0}&,\quad i_c = i_r,\\
        c_{+1}&,\quad i_c = i_r + 1,
    \end{aligned}\right.
\end{equation}
where $i_r, i_c = 0,1,\dots (N_x-1)$, $A_{(N_x-1), N_x} \equiv A_{(N_x-1),0}$, $A_{0, -1} \equiv A_{0,(N_x-1)}$, and the constant scalars are
\begin{equation}
    c_{0} = - \frac{2 D}{\Delta x^2},\quad c_{\pm 1} = \left(\frac{D}{\Delta x^2} \mp \frac{v}{2\Delta x}\right),\quad \Delta x = (N_x - 1)^{-1}.
\end{equation}
For the LCHS simulations, the matrix $A$ is decomposed into Hermitian components $A_L$ and $A_H$ according to Eq.~\eqref{eq:A-decomposition}, and then one solves Eq.~\eqref{eq:psi-LCHS} where the LCHS operator $U_{\rm LCHS}$ is represented by the weighted sum~\eqref{eq:LCHS-discr}.
The block-encoding oracles computing the matrices $A_L$ and $A_H$ are constructed using the general block-encoding technique for sparse matrices described in Refs.~\onlinecite{Novikau22, Novikau24-EVM}.
The corresponding circuits are shown in Fig.~\ref{circ:be-AH-AL}.
There, the register $r_x$ stores the spatial distribution of the variable $\psi$.
The ancillary qubits $a_{x,j}$ are used to address matrix elements either on the main matrix diagonal, or on the matrix left or right sidebands.
The ancilla $a_e$ serves as a target qubit for the rotation operators applied for computing the values of the matrix elements.
Together, the registers $a_x$ and $a_e$ compose the register $a_A$ in Fig.~\ref{circ:be-general}.
In the block-encoding oracle of the matrix $A_H$, we use a combination of $R_y$ and $R_z$ gates to encode complex values:
\begin{equation}\label{eq:Rc}
    R_c(\zeta_y, \zeta_z) \equiv R_y(\zeta_y) R_z(\zeta_z).
\end{equation}
The rotation angles $\zeta$ used in the circuits~\ref{circ:be-AH-AL} are computed in the following way
\begin{subequations}\label{eq:zeta}
\begin{eqnarray}
    &&\zeta_{y,\pm 1} = 2 \arccos{(|A_{H,j,j\pm 1}|/\eta_{H})},\\
    &&\zeta_{z,\pm 1} = - 2 \arg{(|A_{H,j,j\pm 1}|/\eta_{H})},\\
    &&\zeta_l = 2 \arccos{(|A_{L,j,j+l}|/\eta_{L,l})}, \quad l = -1, 0, 1,
\end{eqnarray}
\end{subequations}
where $\eta_{H}$ and $\eta_{L,l}$ are real coefficients used to properly normalize the matrix elements $A_{H,j,j\pm1}$ and $A_{L,j,j+l}$, correspondingly.
The exact values of these coefficients can be found in Ref.~\onlinecite{code-OPT-LCHS}.

The emulation of the resulting LCHS circuit is performed using the QuCF framework \cite{QuCF}.
The total number of qubits in the circuit is 
\begin{equation}
n_{\rm LCHS}=n_x + n_k + n_{\rm BE} + n_{\rm QSP} + n_w + \ycb{n_{{\rm AA}, w}} + n_{\rm init} 
\end{equation}
where 
$n_x = \log_2 N_x = 6$ qubits are used for encoding the variable $\psi(x)$, 
$n_{\rm BE} = 5$ ancillary qubits are used for block-encoding the matrices $C_j$,
$n_{\rm QSP} = 2$ ancillae are used for constructing the QSP circuit,
\ycb{$n_w = 1$ ancilla is used for computing the LCHS weights (as a reminder, as discussed in Sec.~\ref{sec:circuit}, a brute-force direct computation of the LCHS weights is used in these numerical simulations).
Also, $n_{{\rm AA}, w} = 2$ ancillae are used for AA of the LCHS weights,} 
and $n_{\rm init} = 2$ ancillae are used for computing the initial conditions. 
For instance, in the case with $n_k = 12$ discussed below, the total number of qubits in the LCHS circuit is $30$.

\begin{figure}[!t]
\centering
\includegraphics[width=0.49\textwidth]{ 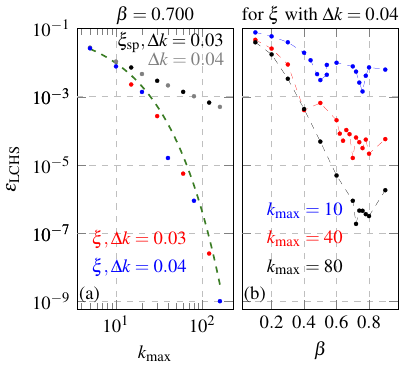}
\caption{
    \label{fig:scan-err-py}
    Results from LCHS simulations of the ADE for $t = 0.8$ without invoking the LCHS circuit, i.e. classical simulations of the LCHS equation~\eqref{eq:psi-LCHS}.
    (a) 
    The dependence of $\yerr$ on $k_{\rm max}$ in the LCHS simulations using the special kernel~\eqref{eq:xi-special} (black and gray markers) and the improved kernel~\eqref{eq:xi-general} with various $\Delta k$ and with $\beta = 0.7$ (colored markers).
    The green dashed line approximates the error with the fitting function~\eqref{eq:err-fitting}.
    (b) 
    The dependence of $\yerr$ on $\beta$ in the LCHS simulations with the improved kernel~\eqref{eq:xi-general} for various $k_{\rm max}$ and for $\Delta k = 0.04$.
}
\end{figure}

First of all, we solve Eq.~\eqref{eq:psi-LCHS} directly without invoking the LCHS circuit~\ref{circ:LCHS}.
The results of these simulations for various LCHS kernels and various values of the scalar $\beta$ are shown in Fig.~\ref{fig:scan-err-py}.
In particular, in Fig.~\ref{fig:scan-err-py}a, one can see there that the scaling of the truncation error $\yerr$ with $k_{\rm max}$ can be exponentially improved by using the kernel~\eqref{eq:xi-general}.
In this case, the scaling can be approximated reasonably well by the function
\begin{equation}\label{eq:err-fitting}
    \yerr \approx 0.12 
    \exp{(-0.5 k_{\rm max}^\beta)},
\end{equation}
which 
confirms the theoretical scaling~\eqref{eq:err-scaling-impr-theory}. 
\ycb{The approximate numerical value observed for the exponent $\approx 0.5$ is close to the bound in Eq.~\eqref{eq:error_trunc} which predicts $\cos{(\beta \pi/2)}\approx 0.454$ for $\beta=0.7$.
Note that, using the bound of Eq.~62 in \onlinecite{An23impr} would have resulted in 1/2 this value, $\cos{(\beta \pi/2)}/2\approx 0.227$, which is clearly ruled out by the numerical data.}
According to Fig.~\ref{fig:scan-err-py}b, 
\ycb{for these choices of parameters, such as $D$ or $v$, the error is minimized by choosing $\beta$} in the interval between $0.7$ and $0.8$, which is consistent with the numerical results of Ref.~\onlinecite{An23impr}.

\begin{figure}[!t]
\centering
\includegraphics[width=0.49\textwidth]{ 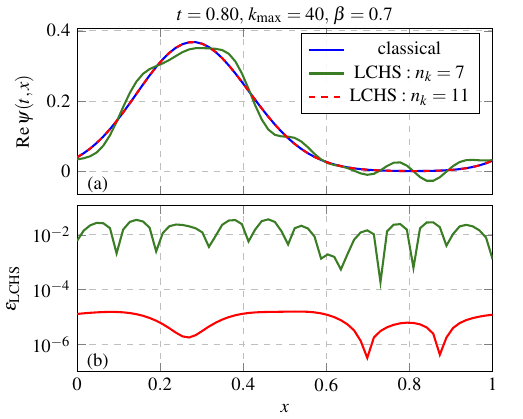}
\caption{
    \label{fig:qucf-signals}
    (a) A comparison between the exact classical simulation (blue line) and the approximate simulations using the LCHS circuit~\ref{circ:LCHS} for various $n_k$ (green and red lines).
    (b) The error of the LCHS simulations.
}
\end{figure}

Now, to test the LCHS method mapped on the LCU circuit~\ref{circ:LCHS} using the coordinate transformation~\eqref{eq:coord-transf} and the kernel~\eqref{eq:xi-general}, we simulate this circuit with different $k_{\rm max}$ and $n_k$.
The circuit is constructed using single-target multicontrolled (STMC) gates such as $X$, $H$, $R_y$, and $R_z$ gates controlled by multiple qubits.
According to Refs.~\onlinecite{Barenco95, Claudon24, Guseynov26}, an STMC gate controlled by $n_c$ qubits can be represented by a circuit with $\oO(n_c)$ elementary gates without using ancillae.

\begin{figure*}[!t]
\centering
\includegraphics[width=0.99\textwidth]{ 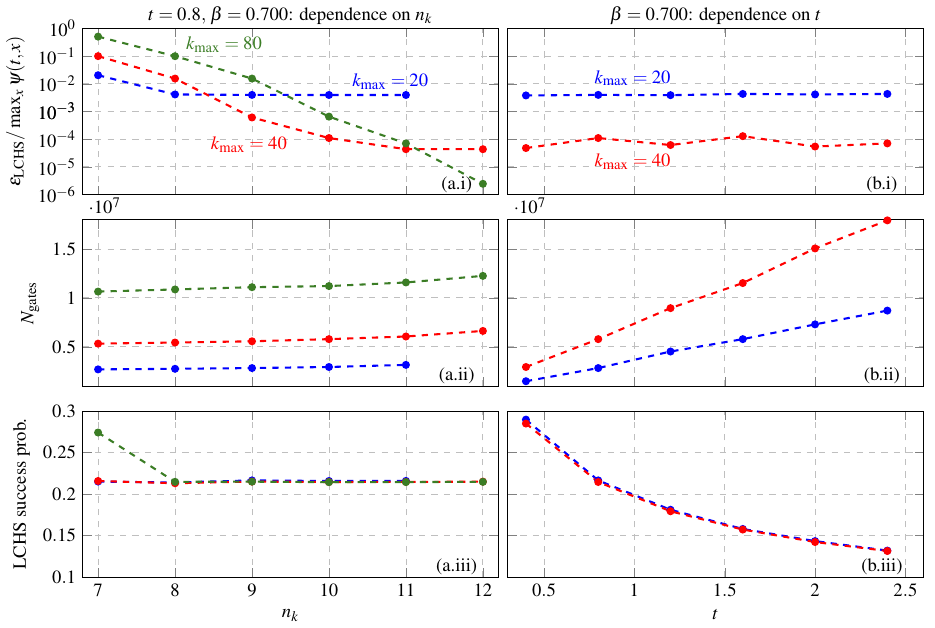}
\caption{
    \label{fig:scan-err-qucf}
    The dependence of the normalized LCHS error (a.i), the number of STMC gates (a.ii), and the success probability of the LCHS circuit (a.iii) on $n_k$ in the LCHS simulations using the circuit~\ref{circ:LCHS} for various $k_{\rm max}$.
    The dependence of the error (b.i), STMC gate count $N_{\rm gates}$ (b.ii), and 
    success probability (b.iii) on $t$ for various $k_{\rm max}$.
    For $k_{\rm max} = 20$, the cases with $t = 0.4$ and $0.8$ have $n_k = 9$, the cases with $t = 1.2$ and $1.6$ have $n_k = 10$, and the cases with $t=2.0$ and $2.4$ have $n_k=11$.  
    For $k_{\rm max} = 40$, the corresponding values of $n_k$ are increased by one.
}
\end{figure*}

The signal $\psi(t,x)$ at $t = 0.8$ and the corresponding error in the LCHS simulation with the circuit~\ref{circ:LCHS} are shown in Fig.~\ref{fig:qucf-signals}.
In particular, one can see that it is possible to decrease the error up to around $10^{-5}$ using $k_{\rm max} = 40$ with $n_k = 11$.
The scaling of the LCHS error, the number of STMC gates in the circuit, and the circuit success probability are shown in Fig.~\ref{fig:scan-err-qucf} as functions of $k_{\rm max}$, $n_k$, and the simulated time interval $t$. 
Clearly, by increasing $k_{\rm max}$ and $n_k$ for a given time instant $t$, one can exponentially increase the LCHS precision.
At the same time, the number of gates in the LCHS circuit increases linearly with $k_{\rm max}$ and only logarithmically with $N_k$, while the success probability of the circuit stays near the same level.
This is consistent with Eqs.~\eqref{eq:Q-be} and~\eqref{eq:sel-scaling-final}.
\ycb{On the other hand, the circuit success probability decreases with time because the simulated signal $\psi(t,x)$ decays in time due to the imposed diffusivity $D$.}
Finally, the number of gates in the circuit grows linearly with time while the LCHS precision stays at nearly the same level.
If we take into account the decomposition of the circuit into elementary operators~\cite{Barenco95, Claudon24, Guseynov26}, the number of elementary gates should scale at least as $\oO(N_{\rm gates} (n_x + n_k))$ where $N_{\rm gates}$ is the number of STMC gates shown in Fig.~\ref{fig:scan-err-qucf}, and we make a conservative estimate that each STMC gate is controlled by at least $n_x + n_k$ qubits.
The actual number of elementary gates will strongly depend on the available set of native gates on a chosen quantum device, as well as on the topology and qubit interconnectivity of the quantum hardware.
\ycb{Here, we point out that, if the size of the ancillary register is increased by $[2-4]\times n_k$, as would occur for the approach of \onlinecite{Low25}, then our estimate for the cost of these STMC gates would need to increase by a similar amount, e.g. if the register is size $3n_k$, then the estimate for the cost of the STMC gates increases to $\oO(N_{\rm gates} (n_x + 3n_k))$. }

Thus, we demonstrated through actual numerical emulation of the LCHS circuit that our block-encoding technique, Fig.~\ref{circ:qubitization}, preserves all theoretical benefits of the LCHS algorithm and simplifies its mapping onto a quantum circuit.
In particular, our numerical simulations prove that the analytical results of the seminal works~\onlinecite{An23, An23impr} about the high success probability of the LCHS algorithm, the exponentially fast error convergence, and the linear scaling with time all hold for our highly efficient explicit implementation.

\section{Conclusions}\label{sec:conclusions}
In this work, an efficient quantum algorithm based on the Linear Combination of Hamiltonian Simulations (LCHS) has been proposed for simulating dissipative initial-value problems.
In this method, a nonunitary operator represented by an exponential function of a non-Hermitian generator is approximated by a weighted sum of Hamiltonian evolutions that depend on the Hermitian components of the generator and on an additional Fourier coordinate.

By recasting the Fourier coordinate as a trigonometric function, we derived a highly efficient encoding of the LCHS sum into a quantum circuit.
\ycb{This enables the use of Fej\'er-Clenshaw-Curtis (FCC) quadrature for LCHS, which is equivalent to Chebyshev-Gauss-Lobatto (CGL) quadrature.
As proven in Theorem~\ref{thm:FCC-LCHS}, this method has the best convergence of all proposed quadrature methods for this purpose.
It also allows one to eliminate trotterization and perform all Hamiltonian simulations with a single QSP circuit.
In turn, this improves the complexity scaling of the selector in Eq.~\ref{eq:sel-scaling-final} to the optimal value of linear in time.
It also significantly improves the complexity of the weight computation in Eq.~\ref{eq:Qw_FCC}, as this now  no longer increases linearly with time as it does in \onlinecite{An23impr,Novikau24KvN}.
These results can be recast in terms of precision in the form of Eq.~\ref{eq:Qsel_FCC_precision} for the selector and Eq.~\ref{eq:Qw_FCC_precision} for the weights.
Note, however, that for dissipative evolution, the decay of the solution in time causes an avoidable need to boost the success probability, and, hence, amplifies the cost by the factor $\|\psi(0)\|/\|\psi(t)\|$ unless one performs a spectral shift to eliminate the decay.
Finally, we proved that this method can also be applied to the recently proposed approximate LCHS methods,\cite{Low25,Jin25} and this would improve the scaling of our algorithm with error from the near-optimal value of $\log^{1/\beta}(\yerr^{-1})$, where $\beta\in[0.7,0.8]$, i.e. $1/\beta\in[1.25,1.43]$, to the optimal value of $\log(\yerr^{-1})$.
Testing the FCC-LCHS method on these new kernels is an important topic for future work.
}

The quantum circuit was tested by \ycr{simulating} the advection-diffusion equation with a uniform velocity and diffusivity, and the numerical simulations confirmed the scaling of the proposed circuit.
The proposed encoding of dissipative problems can be used for solving a wide class of nonunitary differential equations including the Liouville equation and various linear embedding techniques of nonlinear problems.

\ycb{We summarize that, relative to \onlinecite{Low25}, our method achieves benefits in memory and computational cost by eliminating the need for a larger ancillary register of size $[2-4]\times n_k$ and at least $[2-4]\times \oO(3n_k)$ additional 2-qubit gates.
That method also increases the cost estimate for the overall number of elementary gates required to represent the single-target-multicontrolled (STMC) gates in the circuit, again rising with the increase in the size of the ancillary $k$-register.
}

\ycb{It should be straightforward to extend our method to the case of systems with a time-dependent generator and inhomogeneous forcing.
It is also important to understand the numerical convergence of LCHS for more complicated inhomogeneous differential equations with spatially-dependent and time-dependent coefficients.
We plan to explore these interesting questions in future work.
}

\section*{Acknowledgments}
\ycb{We would like to sincerely thank an anonymous referee for asking us to compare our work to that of two manuscripts \onlinecite{Low25,Jin25} that were posted to arXiv after the submission of our work to the journal.  
These works are the first we are aware of to investigate optimal approximate LCHS methods and are based on an analytic continuation approach similar to that first proposed for approximating exponential decay in \onlinecite{Silva2022fourier}.
In particular, our extension of the theorems in \onlinecite{Low25} to near-optimal kernels allowed us to significantly improve our original convergence estimates, based on \onlinecite{An23,An23impr,Novikau24KvN}.
To cleanly separate the priority of the different ideas in this paper, the additional error and complexity analysis that we performed after our original submission is almost entirely presented in Sec.~\ref{sec:error-analysis}. 
}

This work, LLNL-JRNL-2001345, was supported by the U.S. Department of Energy (DOE) Office of Fusion Energy Sciences “Quantum Leap for Fusion Energy Sciences” Project No. FWP-SCW1680 at Lawrence Livermore National Laboratory (LLNL). 
Work was performed under the auspices of the U.S. DOE under LLNL Contract DE-AC52–07NA27344.
This research used resources of the National Energy Research Scientific Computing Center, a DOE Office of Science User Facility supported by the Office of Science of the U.S. Department of Energy under Contract No. DE-AC02-05CH11231 using NERSC award FES-ERCAP0028618.

\appendix

\bibliography{main}

\end{document}